\documentclass{emulateapj}
\usepackage{epsfig}
\usepackage{color}

\def\mbh{M_{\bullet}}
\def\mbhp{M_{\bullet,\rm 1}}
\def\mbhs{M_{\bullet,\rm 2}}
\def\teff{T_{\rm eff}}
\def\rin{r_{\rm in}}
\def\rg{r_{\rm g}}
\def\mp{m_{\rm p}}
\def\sigB{\sigma_{\rm B}}
\def\sigT{\sigma_{\rm T}}
\def\fedd{f_{\rm E}}
\def\re{R_{\rm E}}
\def\hp{h}
\def\abbh{a_{\rm BBH}}
\def\um{\mu \rm m}
\def\msun{M_{\odot}}
\def\kms{\rm km~s^{-1}}

\def\be{\begin{equation}}
\def\ee{\end{equation}}
\def\bea{\begin{eqnarray}}
\def\eea{\end{eqnarray}}

\shorttitle{Microlensing of Sub-parsec Massive Binary Black Holes}
\shortauthors{Yan et al.}

\begin{document}
\title{Microlensing of Sub-parsec Massive Binary Black Holes in
Lensed QSOs: Light Curves and Size-Wavelength Relation}

\author{Chang-Shuo Yan$^1$, Youjun Lu$^1$, Qingjuan Yu$^2$, 
Shude Mao$^1$ and Joachim Wambsganss$^3$}
\affil{$^1$~National Astronomical Observatories, Chinese Academy of
Sciences, Beijing, 100012, China; yancs, luyj@nao.cas.cn \\
$^2$~Kavli Institute for Astronomy and Astrophysics, Peking
University, Beijing, 100871, China \\
$^3$Astronomisches Rechen-Institut am Zentrum f\"{u}r Astronomie der
Universit\"{a}t Heidelberg, M\"{o}nchhofstrasse 12-14,
69120 Heidelberg, Germany 
}

\begin{abstract}

Sub-parsec binary massive black holes (BBHs) are long anticipated to
exist in many QSOs but remain observationally elusive. In this paper,
we propose a novel method to probe sub-parsec BBHs through
microlensing of lensed QSOs. If a QSO hosts a sub-parsec BBH in its
center, it is expected that the BBH is surrounded by a circum-binary
disk, each component of the BBH is surrounded by a small accretion
disk, and a gap is opened by the secondary component in between the
circum-binary disk and the two small disks. Assuming such a BBH
structure, we generate mock microlensing light curves for some QSO
systems that host BBHs with typical physical parameters. We show that
microlensing light curves of a BBH QSO system at the
infrared-optical-UV bands can be significantly different from those of
corresponding QSO system with a single massive black hole (MBH),
mainly because of the existence of the gap and the rotation of the BBH
(and its associated small disks) around the center of mass. We
estimate the half-light radii of the emission region at different
wavelengths from mock light curves and find that the obtained
half-light radius vs. wavelength relations of BBH QSO systems can be
much flatter than those of single MBH QSO systems at a wavelength
range determined by the BBH parameters, such as the total mass, mass
ratio, separation, accretion rates, etc.  The difference is primarily
due to the existence of the gap.  Such unique features on the light
curves and half-light radius-wavelength relations of BBH QSO systems
can be used to select and probe sub-parsec BBHs in a large number of
lensed QSOs to be discovered by current and future surveys, including
the Panoramic Survey Telescope and Rapid Response System (Pan-STARRS),
the Large Synoptic Survey telescope (LSST) and Euclid.

\end{abstract}


\section{Introduction}\label{sec:intro}

Massive binary black holes (BBHs) on sub-parsec scales are unavoidable
products of mergers of galaxies if each galaxy contains a central
massive black hole \citep[MBH; e.g.,][]{BBR80,Yu02, MM05}. In the
$\Lambda$CDM cosmology, big galaxies are formed through hierarchical
mergers of small galaxies, therefore BBHs may exist in the centers of
many galaxies. In gas-rich mergers, the consequent sub-parsec BBHs may
be surrounded by a circum-binary disk, within which a gap is opened by
the inspiraling secondary MBH, and each component of the BBH may be
associated with a small accretion disk within the gap, fed by material
infalling from the inner edge of the circum-binary disk
\citep[e.g.,][]{AL94, Ivanov99, Escala05, Hayasaki07, Hayasaki08,
Dotti07, Cuadra09, DOrazio12, Rafikov12, Farris13}. These active BBHs
may also emerge as QSOs but may have some unique features
distinguishable from single active MBH systems.

Observational search for and identifying BBHs is important because
they are not only probes to the current galaxy formation model but
also main sources for gravitational waves. Recent observations have
already revealed a few active MBH pairs (dubbed as dual AGNs) on kpc
scale through various techniques \citep[e.g.,][]{Komossa03, Wang09,
Comerford09, Liu10, Shen11, Fu11, Rosario11, Yu11}, which are
presumably the precursors of sub-parsec BBHs. However, spatially
resolving and identifying sub-parsec BBHs (both active and inactive
ones) is still beyond the capabilities of the present-day telescopes
\citep[e.g.,][]{Yu02}.  A number of methods have thus been proposed
and applied to probe the BBH systems indirectly. Such methods include
the double-peaked or asymmetric broad emission line
\citep[e.g.,][]{BL09, Tsalmantza11, Eracleous12, Ju13, Shen13}, flux
ratio of different emission lines \citep{Montuori11, Montuori12},
periodic variation of the QSO light curves
\citep[e.g.,][]{Valtonen08, Sesana12, GM12, Hayasaki12}, double broad
relativistic Fe K$\alpha$ lines \citep[e.g.,][]{YL01, Sesana12,
McKernan13}, and kinematic signature \citep{Meiron13}, etc. Most of
these methods depend on the orbital motions of BBHs. A number of
sub-parsec BBH candidates have been reported but few of them were
confirmed because of various complications in these methods, which
make the differences of the properties of BBH systems from those of
its alternatives ambiguous.  It is therefore imperative to devise new
method/approach to unambiguously reveal sub-parsec BBHs/binary QSOs.

It has been demonstrated that the microlensing magnification of lensed
QSOs \citep[e.g.,][]{WPS90,WMS95} is a powerful tool to probe the size
and even temperature structure of accretion disks, as the Einstein
radius of a single star in the foreground lens galaxies is comparable
to the disk size of background QSOs and the QSO flux variation is
determined by the ratio of the Einstein radius to the size of the
emission region \citep[e.g.,][]{Wambsganss06}.  This method has been
applied to estimate the disk sizes for more than a dozen lensed QSOs,
and the results show that the estimated disk temperature profile is
consistent with the thin disk model though the estimated disk size may
be a factor of $\sim 4-5$ larger than the standard thin disk
prediction \citep[e.g.,][]{WP91, Mortonson05, Eigenbrod08, Poindexter,
Morgan10, Blackburne11,JV12}.  \citet{DA11} and \citet{AS12} suggest
that alternative disk structure models, such as inhomogeneous disk or
super-Eddington accretion disk, may be required to explain the
microlensing results. Long term monitoring of microlensing of lensed
QSOs are helpful to identify the detailed structure of accretion disks
associated with these lensed QSOs. 

The structure of disk accretion onto BBHs is fundamentally different
from that onto a single MBH. The most distinct feature of an active
BBH system should be the gap opened by the inspiraling secondary MBH
\citep[e.g.,][]{AL94, Ivanov99, Dotti07, Hayasaki07, Hayasaki08,
Cuadra09}.  Direct detection of such a feature is not possible
currently. In this paper, we propose a novel method to probe
sub-parsec BBHs/binary QSOs by detecting signatures of the gap through
the microlensing of lensed QSOs. 

This paper is organized as follows. In section~\ref{sec:config}, we
first illustrate the configuration of an active BBH system and
calculate the surface brightness distribution of such a system at the
given UV-optical-infrared bands by adopting a simple thin disk
accretion model. In section~\ref{sec:mocklc}, mock microlensing light
curves are generated by convolving the brightness distribution of
background sources, either an active MBH or an active BBH system, with
the foreground magnification maps. Adopting the Monte-Carlo
microlensing analysis method developed by \citet{Kochanek04}, we
demonstrate that the size of emission region can be estimated for a
BBH system through the mock light curves by assuming a single standard
accretion disk model in section~\ref{sec:disksize}. As shown in
section~\ref{subsec:results}, the expected relationship between disk
size and wavelength for a BBH system show some apparent anomaly at
some wavelength compared with that for a single MBH system. We propose
that this anomaly can be taken as a signature of BBHs and should be
useful for selecting BBH candidates from lensed QSOs with current and
future facilities.  Discussions and conclusions  are given in
sections~\ref{sec:complications} and \ref{sec:discon}.

\section{Configuration of accretion onto massive binary black
holes}\label{sec:config}

In order to calculate the microlensing signature of BBH systems, it is
necessary to first know the structure of accretion onto these BBHs. In
this section, we adopt a simple disk accretion model for BBH systems
and then calculate the brightness distribution of these BBH sources.
By convolving the background source brightness distributions with
foreground magnification maps due to lens galaxies, we then obtain
mock light curves for these sources.

\begin{figure}
\centering
\includegraphics[scale=0.45,angle=-90]{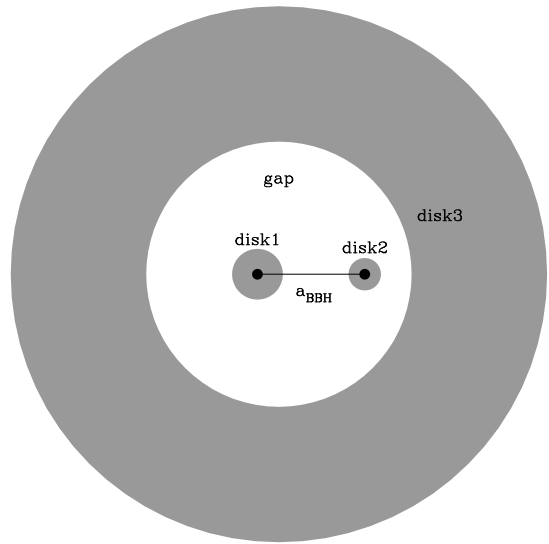}

\caption{Schematic diagram for a sub-parsec BBH system. The BBH system is
assumed to be on a circular orbit with semimajor axis of $a_{\rm
BBH}$. The two MBHs are surrounded by a circum-binary disk (disk3),
which is truncated at the edge of an inner gap opened by the secondary
MBH. Within the gap, each component of the BBH also has its own small
accretion disk truncated at half of the mean Roche radius (disk1 and
disk2).  }

\label{fig:f1}
\end{figure}

\subsection{Geometry}

Considering a BBH system, the semimajor axis of the BBH is $\abbh$ and
the masses of the primary and secondary MBHs are $M_{\bullet,1}$ and
$M_{\bullet,2}$, respectively. As revealed by a number of numerical
simulations \citep[e.g.,][]{AL94, Ivanov99, Dotti07, Hayasaki07,
Hayasaki08, Cuadra09, Farris13}, if initially the primary MBH is
surrounded by a massive disk, the secondary MBH may quickly settle
down to the disk and open a gap in the disk with width of $\sim
2R_{\rm RL}$, where $R_{\rm RL}$ is the Roche radius and  roughly
given by $\abbh (M_{\bullet,2} /3 M_{\bullet,1})^{1/3}
=\abbh(q/3)^{1/3}$, and $q$ is the ratio of the secondary MBH mass to
the primary MBH mass. A circum-binary disk still exists outside of the gap;
within the gap each MBH may be associated with its own disk. We
simply assume the outer boundary of the accretion disk associated with
each MBH is determined by the mean Roche radius \citep{Eggleton}
\be
R_{\rm RL}(x)=0.49\abbh x^{2/3}/[0.6x^{2/3}+\ln(1+x^{1/2})],
\label{eq:roche}
\ee
where $x$ is the mass ratio of the disk MBH to the other MBH and it is
$q$ and $1/q$ for the secondary and primary MBH, respectively.
Considering that each disk may not fill the whole Roche lobe as
indicated by numerical simulations, we assume the disk size is about
half of the mean Roche radius.\footnote{If assuming the disk size is
the same as the Roche radius, we find that our results on the half-light
radius-wavelength relation are not significantly changed qualitatively.} 
Figure~\ref{fig:f1} shows a schematic diagram of such a
system with $q=1/4$. For simplicity, we only consider those cases where
BBHs are on circular orbits and their associated triple disks are
coplanar. More complicated cases, such as eccentric BBHs and
non-coplanar disks, etc., will be further discussed in
section~\ref{sec:complications}.

\subsection{A simple model for accretion onto MBH/BBH systems}
\label{subsec:diskmodel}

\subsubsection{Thin disk model for accretion onto a single
MBH}\label{subsec:sd}

For simplicity, we adopt the standard thin disk model to describe the
radiatively efficient QSO accretion process around a single MBH
\citep[e.g.][]{SS73,NT73}. In such a model, the emission from an
annulus $r- dr/2\to r+dr/2$ of the disk is approximated by black body
radiation with effective temperature of
\bea
\teff(r) & = & \left[\frac{3G\mbh \dot{M}_{\rm acc}}{8\pi\sigB
r^3}\left(1-\sqrt{\frac{\rin}{r}}\right)\right]^{1/4} \nonumber \\
& \simeq & 2\times 10^5~{\rm K} \left(\frac{0.1}{\epsilon}
\right)^{1/4} \left(\frac{f_{\rm E}}{0.3}\right)^{1/4} 
\left(\frac{10^8\msun}{M_{\bullet}}\right)^{1/4} \nonumber \\
& & \times \left(\frac{\rin}{r}\right)^{3/4}
\left(1-\sqrt{\frac{\rin}{r}}\right)^{1/4},
\label{eq:temp}
\eea
where $G$ is the gravitational constant, $\sigB$ is the
Stefan-Boltzmann constant, $\mbh$ and $\dot{M}_{\rm acc}$ are the mass
and accretion rate of the MBH, $\rin$ is the radius of the disk inner
edge, $\epsilon$ is the radiative efficiency, and $\fedd$ is the
Eddington ratio. The Eddington ratio here is defined as
$\fedd=\dot{M}_{\rm acc}/ \dot{M}_{\bullet,\rm E}$, where
$\dot{M}_{\bullet,\rm E}= 4\pi\mp G \mbh /\epsilon c \sigT$ is the
Eddington accretion rate, $c$ is the speed of light, $\mp$ is the proton
mass, and $\sigT$ is the Thompson scattering cross section. The
infrared-optical-UV radiations concerned in the following calculation
comes from disk regions faraway from the MBH event horizon, the
general relativistic effects due to the central MBH and its spin are
not significant. Observations on QSOs and MBHs have shown that
$\epsilon \sim 0.1$ \citep[e.g.,][]{YT02, Marconi04,YL04, YL08,
Shankar09, Shankar13, Wu13}, which corresponds to $\rin \sim 
3.5 \rg$ if adopting the standard thin accretion disk model for
Kerr black holes, where $\rg \equiv GM_{\bullet}/c^2$ is the gravitational
radius. Therefore, we adopt $\rin=3.5\rg$ and ignore the
general relativistic corrections to the temperature profile in the
innermost disk region in this study.  Our results are not
significantly affected if adopting the relativistic thin disk model
by \citet{NT73}.

\begin{deluxetable}{ccccccc}
\tablecaption{Parameters for different systems}
\tablehead{\colhead{Model} & $M_{\bullet}$ & \colhead{q}
& \colhead{$f_{\rm E,1}$} & \colhead{$f_{\rm E,2}$} &
\colhead{$f_{\rm E,c}$} & \colhead{$\abbh(\rg)$} }
\startdata
S0 & $10^8\msun$ & $\cdots$ & $\cdots$  & $\cdots$  & 0.3   & $\cdots$ \\
B1 & $10^8\msun$ &     0.25 & 0.3       & 0.01      & 0.242 & 500 \\
B2 & $10^8\msun$ &     0.25 & 0.3       & 0.01      & 0.242 & 1000 \\
B3 & $10^8\msun$ &     0.25 & 0.3       & 0.01      & 0.242 & 2000 \\
B4 & $10^8\msun$ &     0.25 & 0.3       & 0.01      & 0.242 & 3000 \\
B5 & $10^8\msun$ &     0.25 & $10^{-4}$ & 0.3       & 0.06  & 500 \\
B6 & $10^8\msun$ &     0.25 & 0.3       & 0.3       & 0.3   & 500 \\
B7 & $10^8\msun$ &      1.0 & 0.3       & 0.3       & 0.3   & 500 \\
B8 & $10^8\msun$ &      0.1 & $10^{-4}$ & 0.3       & 0.027 & 500
\enddata

\tablecomments{Basic parameters that define the example single/triple
disk systems. For the S0 system, $M_{\bullet}$ and $f_{\rm E,c}$
represent the mass of the central MBH and Eddington ratio of the
accretion disk, respectively; for other BBH systems, $M_{\bullet}$,
$q$, $a_{\rm BBH}$, $f_{\rm E,1}$, $f_{\rm E,2}$, and $f_{\rm E,c}$
represent the total mass, the mass ratio, the semimajor axis, the
Eddington ratios of the disks associated with the primary component,
secondary component, and the circum-binary disk of each BBHs,
respectively.}

\label{tab:t1}
\end{deluxetable}

According to Equation~(\ref{eq:temp}), the monochromatic specific
intensity at wavelength $\lambda$, as a function of radius $r$, is
given by
\be
B_{\lambda}(r)=\frac{2 hc^2/\lambda^5}{\exp[{\hp c/\lambda k_{\rm
B}\teff(r)}]-1},
\label{bb1}
\ee
where $\hp$ is the Planck constant and $k_{\rm B}$ is the Boltzmann
constant.  For a narrow band filter with central wavelength at
$\lambda$, the half-light radius of the disk ($r_{1/2}$), within which
half of the light is contained, can
be estimated by integrating the specific intensity at $\lambda$ over
the disk radii, and it is given by 
\bea
r_{1/2} &=& 2.4 \cos^{1/2} i
\left[\frac{45 G^2 \mbh^2 m_p \fedd\lambda^4}
{4 \pi^5 \hp c^3 \sigT \epsilon}\right]^{1/3} \nonumber \\
&\simeq &5.1\times 10^{15} \mathrm{cm}
\left(\frac{\mbh}{10^8 \msun}\right)^{2/3}
\left(\frac{\fedd}{\epsilon}\right)^{1/3}
\left(\frac{\lambda}{\um}\right)^{4/3} \nonumber \\
& \simeq & 340 \rg \left(\frac{\mbh}{10^8 \msun}\right)^{-1/3}
\left(\frac{\fedd}{\epsilon}\right)^{1/3}
\left(\frac{\lambda}{\um}\right)^{4/3}, \nonumber \\
\label{eq:rhalf}
\eea
where $i$ is the inclination angle of the disk to the line of sight.
The factor $\cos^{1/2}i$ is included to account for the inclination of
the disk to our line of sight. For simplicity, we set all the disks to
be face on so that $\cos i=1$ ($i=0\arcdeg$) in this study. The half-light
radius scales with the rest frame wavelength as $\propto
\lambda^{4/3}$ and MBH mass as $\propto M_{\bullet}^{2/3}$.

\subsubsection{Simple model for accretion onto massive BBH systems}

For each of the three disks associated with a BBH system (see Figure
1), we assume its temperature profile still follows that given by the
standard thin disk model (eq. 2), except that there is a truncation at
either inner or outer disk. According to the assumptions made in
section 2.2.1, the disks associated with the primary and the secondary
MBHs are truncated at an outer radius of $\frac{1}{2}R_{\rm RL}(1/q)$
and $\frac{1}{2} R_{\rm RL}(q)$, and their inner edges are
$3.5GM_{\bullet,1}/c^2$ and $3.5GM_{\bullet,2}/c^2$, respectively. The
circum-binary disk is truncated at an inner radius of $a_{\rm
BBH}/(1+q) + R_{\rm RL}(q)$. For the circum-binary disk, the mass of
the central accretor is assumed to be the sum of the two BBH
components, i.e., $M_{\bullet,1}+M_{\bullet,2}$.

\begin{figure*}
\centering
\includegraphics[scale=1.0,width=6.0cm,angle=-90]{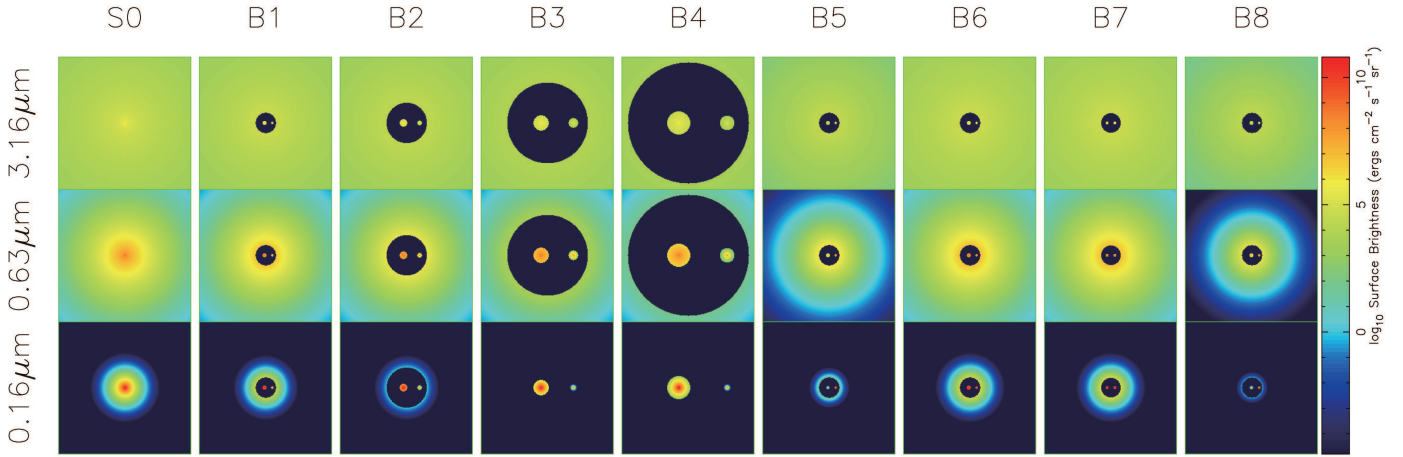}

\caption{Surface brightness distribution maps of disk accretion
systems at a given time. These maps are obtained from the simple disk
models illustrated in section \ref{sec:config}. Dark blue color
represents regions with smaller surface brightness while red bright
color represents regions with larger surface brightness, as indicated
by the side bar. The nine columns, from left to right, show the
results for the system S0, B1, B2, B3, B4, B5, B6, B7, B8,
respectively, and the parameters that define these systems are listed
in Table~\ref{tab:t1}.  The top, middle and bottom rows show the
surface brightness distributions at wavelength $3.16\um$, $0.63\um$,
and $0.16\um$, respectively. Only the central parts of the maps of
each source are shown here, respectively. The side length of
each map shown in this Figure is $4$ times of the Einstein radius
($\re$), and each map has $204$ by $204$ pixels.  }                    

\label{fig:f2}                                                
\end{figure*}

Adopting the above single/triple-disk model, the brightness
distribution of an active MBH/BBH system can be obtained.
Figure~\ref{fig:f2} shows the brightness distribution maps at a given
time and three different wavelengths, i.e., $0.16\um$, $0.63\um$ and
$3.16\um$, for one single MBH system and eight BBH systems (denoted as
S0, B1, B2, B3, B4, B5, B6, B7, B8), respectively. Table~\ref{tab:t1}
lists the basic parameters that define these systems, such as the BBH
semimajor axis ($a_{\rm BBH}$), the total mass ($M_{\bullet}$) and
mass ratio ($q$) of the BBH, and the Eddington ratio of the disk
accretion onto each MBH (i.e., $f_{\rm E,1}$, $f_{\rm E,2}$, $f_{\rm
E,c}=f_{\rm E,1}/(1+q)+qf_{\rm E,2}/(1+q)$), etc.  As seen from
Figure~\ref{fig:f2}, the larger the semimajor axis of a BBH, the
larger the size of the inner disks (disk1 and disk2) and the larger
the inner edge of the circum-binary disk (disk3). For the surface
brightness maps of those BBH systems, the part within the gap should
rotate with time because of the periodic motions of the BBH systems. 

\begin{figure}
\centering
\includegraphics[scale=1.0,width=12.0cm,angle=-90]{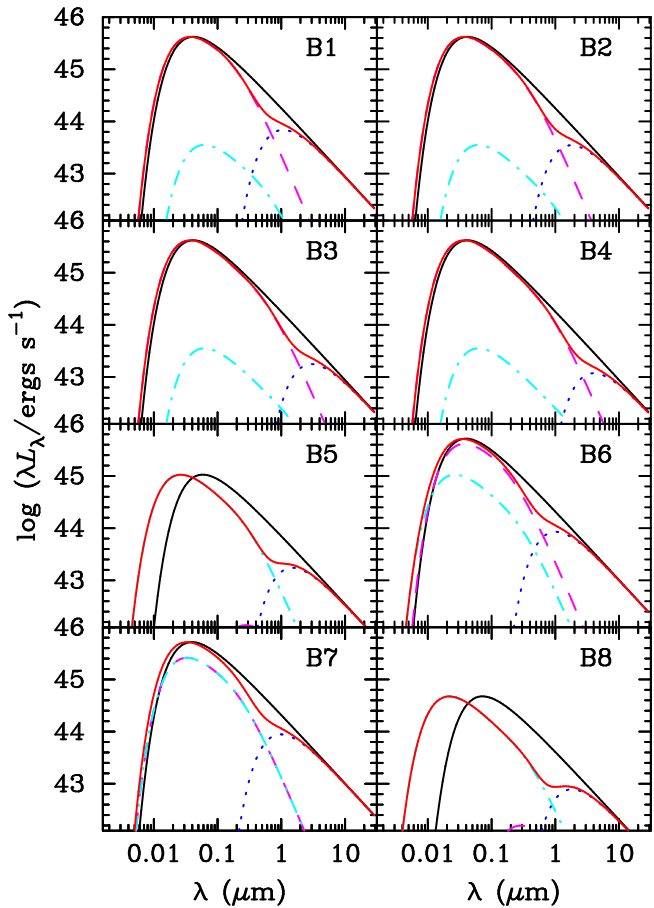}

\caption{Spectral energy distributions for the BBH systems listed in
Table~\ref{tab:t1}. In each panel, the magenta dashed, cyan
dot-dashed, blue dotted lines represent the emission from the primary,
secondary and circum-binary disk, respectively. The total emission
from each BBH system is represented by the red solid line in each
panel. For comparison, the black solid line shows the emission from a
single disk system, of which the MBH mass is the total mass of the two
MBHs and the accretion rate is the same as that of the circum-binary
disk in the corresponding BBH system.  }

\label{fig:f3}
\end{figure}

In the standard thin disk accretion model, the short wavelength
photons come mainly from the inner disk region while the long wavelength
photons from the outer disk region. In the case of a BBH system, a
gap is opened in the primary disk by the secondary MBH (see
Figure~\ref{fig:f1}). Comparing with the radiation from a single MBH
system, therefore, there should be some deficiency of the radiation of
photons from a BBH system in a certain wavelength range, determined by
the inner and outer boundaries of the gap. Figure~\ref{fig:f3} shows the
spectral energy distributions (SEDs) for all those BBH systems listed
in Table~\ref{tab:t1} (i.e., B1, B2, B3, B4, B5, B6, B7, and B8),
according to the simple single/triple-disk model described above. For
those BBH systems shown in Figure~\ref{fig:f3}, we can clearly see the
deficits of radiation in the optical to near infrared bands, 
relative to the corresponding single disk systems. The deficits of
thermal emission in a certain wavelength range have been proposed by
\citet{GM12} as a diagnostic of the BBH systems. The gap in a BBH
system, and correspondingly the deficit of radiation in the gap
region, could result in an anomaly in the half-light radius-wavelength
relation (hereafter denoted as the $r_{1/2}-\lambda$ relation), which
may be significantly different from the smooth $r_{1/2}-\lambda$
relation obtained for the standard thin accretion disk. This
difference may offer an important way to probe BBHs among the lensed
QSOs. 

In order to demonstrate that microlensing light curves and the
$r_{1/2}-\lambda$ relation are useful for selecting/probing BBH
systems, if any, in the distant lensed QSOs, we first generate mock
light curves for a number of assumed BBH systems in
section~\ref{sec:mocklc}, and then use these mock light curves to
extract the $r_{1/2}-\lambda$ relation and compare it with that
obtained for the corresponding single disk systems to reveal the
signature of BBHs in section~\ref{sec:disksize}.

\section{Mock Light curves}\label{sec:mocklc}

The variation of a macroimage magnitude of a lensed point source due
to the microlensing effect can be represented by a magnification map
projected to the source plane, of which the value of each pixel
represents the difference between the macroimage's magnitude measured
by a distant observer when the source at a particular point in the map
and the average macroimage magnitude. This original magnification map
is determined by the properties of the lens (including the surface
stellar mass density and the shear ) and also the distances of the
source and the lens to the observer \citep[e.g.,][]{Kochanek04}. 

We generate the original magnification maps in the source planes by
using the ray-shooting method \citep[e.g.,][]{Wambsganss06, Kayser86,
Paczynski86, Wambsganss90, WPS90}. For simplicity and demonstration purpose,
we adopt fixed values for the mean convergence (surface mass density;
$\kappa$) and shear $(\gamma)$ for all the systems listed in
Table~\ref{tab:t1}, and the values of $(\kappa, \gamma)$ are
arbitrarily set to be $(0.394, 0.395)$, $(0.375, 0.390)$, $(0.743,
0.733)$ and $(0.635, 0.623)$ for the image A, B, C, and D of each
system, respectively, similar to the case for Q$2237+0305$ in
\citet{Kochanek04}. With these settings, images A and B have positive
parity, while images C and D have negative parity. We set the fraction
contributed by stars to the convergence as $\kappa_*/\kappa=0.25$,
where $\kappa_*$ represents the stellar surface density.  The
positions of stars are randomly generated in the model. In principle,
the mass of each star ($m_*$) could be randomly drawn from a
distribution function, e.g., the Salpeter initial mass function
$\propto m_{*}^{-2.35}$, over a finite mass range $m_{*,1}< m_* <
m_{*,2}$. For simplicity and demonstration purpose, we set
$m_*=\left<m_*\right> =m_{*,1} = m_{*,2} =0.3\msun$, where $\left<
m_*\right>$ is the mean mass of the stars. The Einstein radius of a
star in the source plane is fixed to be $\left< \re \right> = 3.0
\times 10^{16} \left(\left<  m_* \right> / 0.3 \msun \right)^{1/2}$cm,
which is typical for the lensed QSOs listed in \citet{Mosquera11}. The
side length of the calculated magnification maps is $40\left<
\re \right>$, and these magnification maps are stored in
$2048\times 2048$ arrays with a pixel size of $\sim 0.02\left< \re
\right>$. 

As lensed QSOs are diffuse sources on the accretion disk scales, it is
necessary to obtain convolved magnification maps by convolving the
original magnification maps with the surface brightness distribution
maps of the sources. The rotation of the BBHs introduces further
complications to the calculation of the convolved magnification maps.
The orbital period of a BBH system is $t_{\rm orb}=2\pi (a^3_{\rm BBH}
/GM_{\bullet})^{1/2}=3.1(M_{\bullet}/10^8\msun) (a_{\rm BBH} /10^3
\rg)^{3/2}$yr. The timescale of the variations due to the microlensing
is determined by the source length scale and the effective relative
moving velocity $v_{\rm e}$ of the source with respect to the lens. In
the following analysis, the light curves of mock lensed QSOs are
``monitored'' over a period of about $\left<\re \right> /
v_{\rm e}$, where $v_{\rm e}=\tilde{v}_{\rm e} (\left< m_* \right> /
0.3 \msun)^{1/2}$, $\hat{v}_{\rm e}$ is the scaled velocity of the
source relative to the lens. The source surface brightness map of
a BBH system may rotate significantly within the monitoring period
depending on the value of $v_{\rm e}$. If $v_{\rm e} \ga 10^4\kms$,
the microlensing caustic passes through the BBH system in a very short
time period $\sim \abbh/v_{\rm e} \la 0.5~{\rm yr} (\abbh/10^3\rg)
(\mbh/10^8\msun)\ll t_{\rm orb}$ for $\mbh\ga 10^8\msun$ and $\abbh
\sim 10^3\rg$, and thus the rotation of the BBH has little effect on
the microlensing light curve within the monitoring period; however, if
$v_{\rm e} \la 1000\kms$, the rotation of the BBH affects the light
curve significantly. For those currently known lensed QSOs, some of
them may have large $v_{\rm e}$ \citep[e.g.,][]{Kochanek04} and some
may have small $v_{\rm e}$ \citep[see][]{Mosquera11}. Therefore, we
consider two cases below: (1) static surface brightness distribution
of the BBH sources, in which we ignore the rotation of the BBH systems
during the monitoring time period; (2) rotating surface brightness
distribution of the BBH sources, in which the rotation of the BBH
systems is simultaneously considered for a typical value of $v_{\rm e}
\simeq 10^3\kms$.

\begin{figure*}
\centering
\includegraphics[scale=0.5,width=6.5cm,angle=-90]{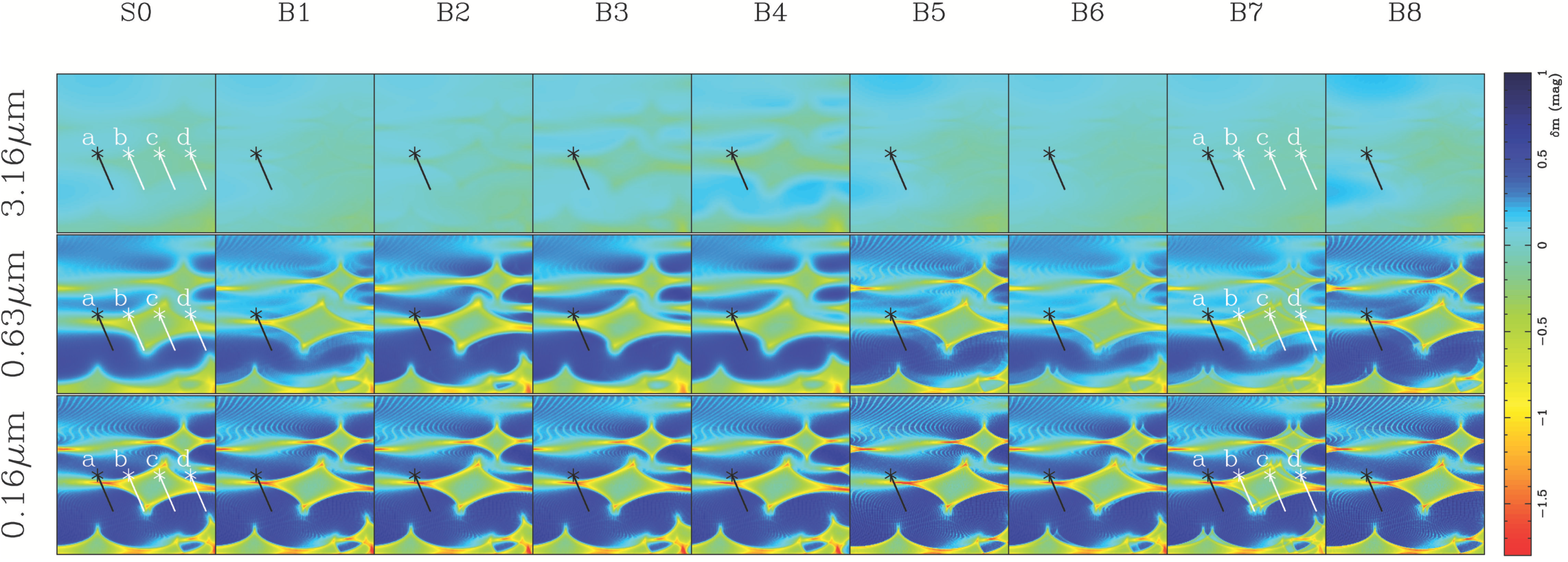}

\caption{Convolved magnification maps in the source planes obtained by
assuming static surface brightness distributions of the sources.
Columns from left to right show the convolved magnification map for
the image A of the system S0, B1, B2, B3, B4, B5, B6, B7, and B8,
respectively (assuming the convergence and shear the same as those for
the lensed QSO 2237+0305). Red color represents regions with high
magnification and consequently small $\delta m$ ($=-2.5\log f$, see
Equation~\ref{Eq:deltam}), while blue colors represent regions with
low magnification, as indicated by the side bar. The top, middle and
bottom rows show the maps at wavelength $3.16\um$, $0.63\um$, and
$0.16\um$, respectively.  The source map of each convolved
magnification map is correspondingly shown in Figure~\ref{fig:f2}. The
black solid lines (denoted as trajectory (a) in each map) and the
white lines (denoted as (b), (c) and (d) in the maps for S0 and B7)
show the trajectories that are adopted in calculating the mock light
curves, and each asterisk symbol represents the starting point of a
trajectory. Only the central parts of the convolution maps for each
system are shown here, respectively.  The side length of each
map shown in this Figure is $4$ times of the Einstein radius in the
source plane, and the map has 204 by 204 pixels.  }

\label{fig:f4}
\end{figure*}

\begin{figure*}
\centering
\includegraphics[scale=1.0,width=13.0cm,angle=-90]{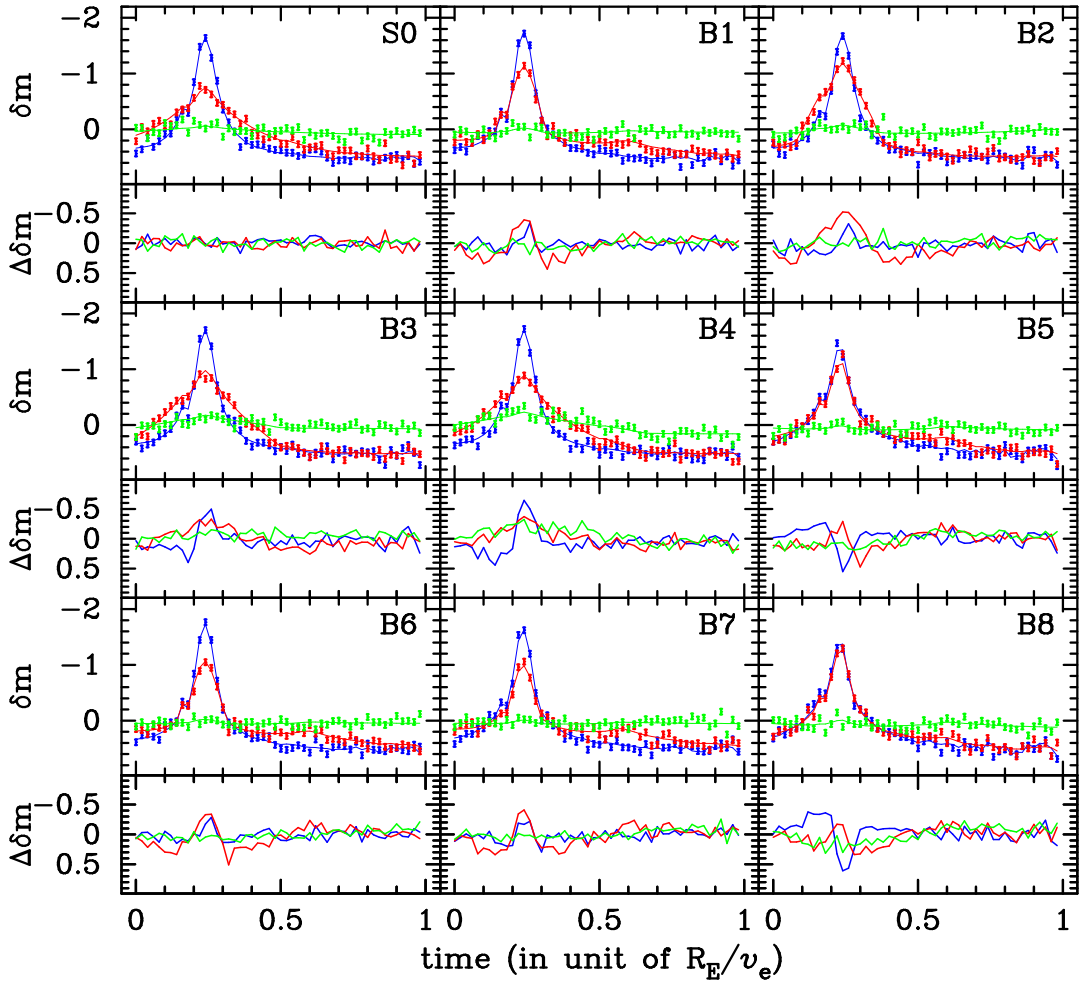}

\caption{Mock light curves obtained for different systems by assuming
static surface brightness distributions of the sources. Panels with
labels show the mock light curves for the image A of the system S0,
B1, B2, B3, B4, B5, B6, B7, and B8, respectively;  a smaller panel
below shows the differences between
the mock light curves of the labeled system and that for a
corresponding single MBH system (with the same mass as the total
mass of the BBH and accretion rate the same as that of the
circum-binary disk in the BBH system). In each panel, points with
errorbars in blue, red and green are for the mock light
curves at wavelength $0.16$, $0.63$ and $3.16\um$, respectively,
by considering the measurement errors; the solid curves in blue,
red and green are for the intrinsic mock light curves at wavelength $0.16$,
$0.63$ and $3.16\um$, respectively, without considering the
measurement errors.  }

\label{fig:f5}
\end{figure*}

\begin{figure*}
\centering
\includegraphics[scale=1.0,width=9.0cm,angle=-90]{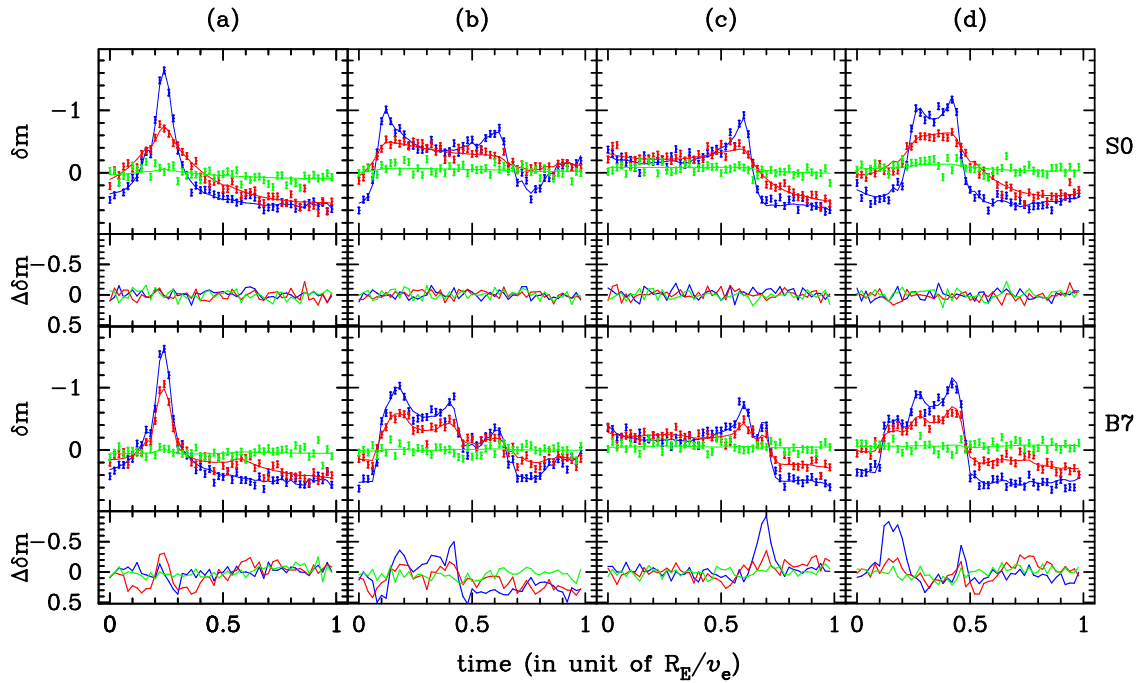}

\caption{Mock light curves obtained for the S0 and B7 systems
shown in Figure~\ref{fig:f4}. Top panels show the mock light curves
obtained for the trajectories (a), (b), (c) and (d) marked for the S0
system in Figure~\ref{fig:f4}, respectively. Bottom panels show the
mock light curves obtained for the trajectories (a), (b), (c) and (d)
marked for the B7 system in Figure~\ref{fig:f4} from left to right,
respectively. Symbols and lines are the same as those in
Figure~\ref{fig:f5}. }

\label{fig:f6}
\end{figure*}

\begin{figure*}
\centering
\includegraphics[scale=1.0,width=13.0cm,angle=-90]{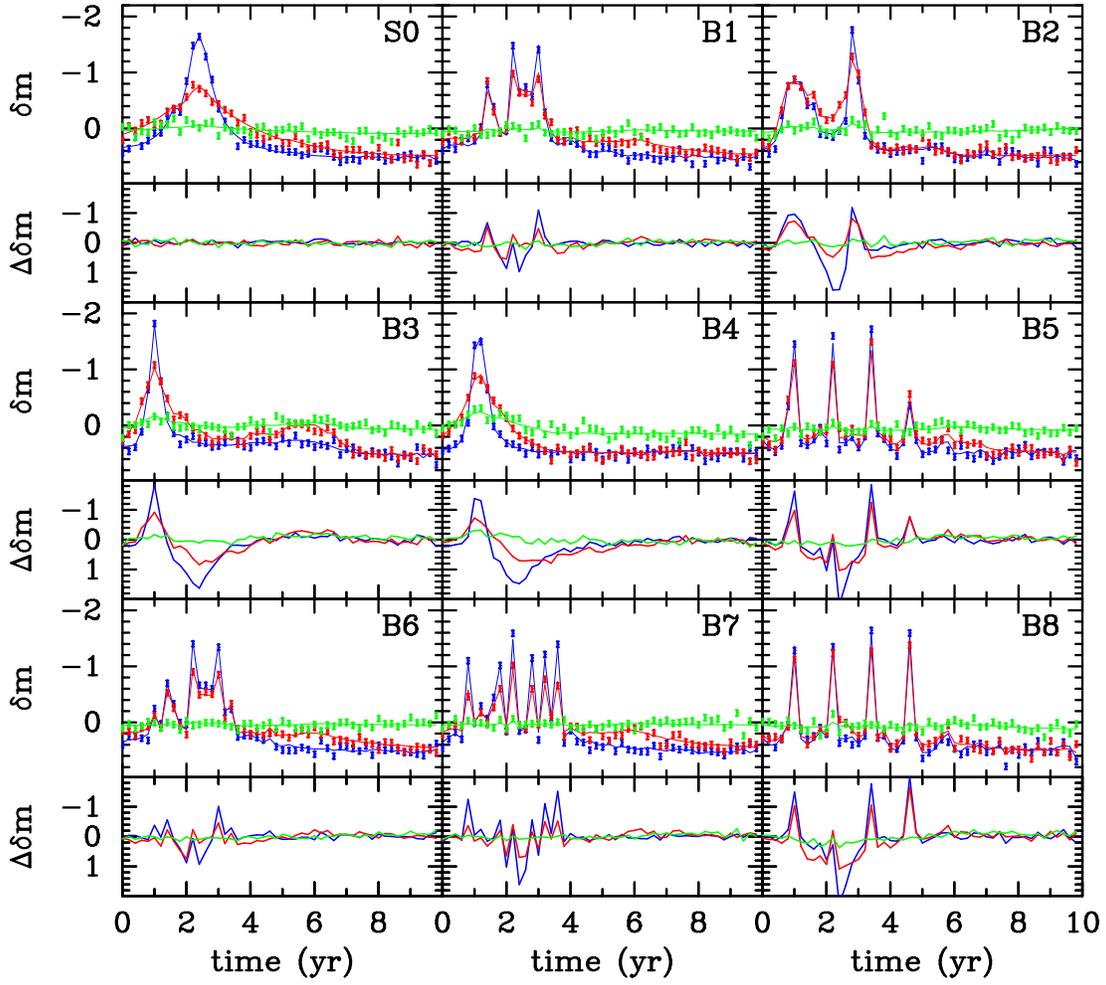}

\caption{Mock light curves obtained for different systems by
considering the BBH rotation. Legends are similar to those for
Figure~\ref{fig:f5}.  }

\label{fig:f7}
\end{figure*}

\subsection{The case of static surface brightness distributions}
\label{subsec:nonrotating}

Assuming that the surface brightness distributions of the sources are
static, we generate the convolved magnification maps in the source
planes for all the systems listed in Table~\ref{tab:t1} by convolving
the original magnification maps with the static surface brightness
maps shown in Figure~\ref{fig:f2}.  Figure~\ref{fig:f4} shows the
central part of each convolved magnification map for the image A of
each system (see Table~\ref{tab:t1}) at wavelength $3.16$ (top
panels), $0.63$ (middle panels), and $0.16\um$ (bottom panels),
respectively.

We assign a trajectory for each image (A, or B, or C, or D) of each
source system listed in Table~\ref{tab:t1} (e.g., the black line,
representing the trajectory (a) in each panel of Figure~\ref{fig:f4}
for image A). For different systems, we assign the same trajectory
with respect to the disk (disk1) associated with the primary MBH.
The length of the trajectory is set to be one Einstein radius. The
effective velocity of the source with respect to the lens is set to
be $v_{\rm e} = 10^4\kms$.  The length of an assigned trajectory thus
corresponds to a period of $\Delta t = t_{\rm E}=\re/v_{\rm e} \simeq
1$~yr.  According to these trajectories, we can calculate the light
curves for each image at each given wavelength (i.e., $\lambda=0.16,
0.25, 0.40, 0.63, 1.0, 1.78, 3.16\um$). We record/measure each light
curve for $50$ times uniformly distributed over the $1$~yr period. We
add a random error to each ``measurement'' by assuming that this error
follows a Gaussian distribution with standard deviation of $0.05$~mag
\citep[see][]{Kochanek04}. Note that the ``measurement'' error may
scale with the magnitude according to the Poisson noise, but this
scaling is ignored in this paper for simplicity. The background
sources could also have some intrinsic variations, and normally the
intrinsic emission of a QSO can vary on a timescale of within a day
\citep[see also][]{Kochanek04}. As the light curve is recorded every
$7$~days, substantially longer than a day, the intrinsic variation of
the sources should appear as incoherent and can be treated as a random
error.  We also add the intrinsic variation to each ``measurement''
according to a Gaussian distribution with standard deviation of
$0.05$~mag \citep[see also][]{Kochanek04}.  Then the ``measured''
magnitude at each epoch $i$ is given by 
\be
m_i=-2.5\log f_i -2.5\log f_{\rm m} + m_{\rm int}+(\delta m|\sigma_1) 
+ (\delta m|\sigma_2), 
\ee
where $f_i$ is the mean amplification value due to microlensing at
epoch $i$, $f_{\rm m}$ is the amplification value due to macrolensing,
$m_{\rm int}$ is the mean intrinsic magnitude, $(\delta m|\sigma_1)$
and $(\delta m|\sigma_2)$ are two quantities to represent the
intrinsic variation and the measurement error, randomly set by
Gaussian distributions with deviations of $\sigma_1 =0.05$ and
$\sigma_2=0.05$, respectively. 

Figure~\ref{fig:f5} shows the resulting mock light curves for the
trajectory (a) in the image A of each system listed in
Table~\ref{tab:t1} at three different wavelength, i.e., $\lambda =
0.16$, $0.63$, and $3.16\um$, respectively.  These light curves
show the deviations of the measured magnitude $m_i$ at each epoch $i$
from the mean magnitude $\left<m_i\right>$, i.e., 
\be
\delta m_i=m_i-\left<m_i\right>= - 2.5\log \left(f_i/\bar{f}\right)
+(\delta m|\sigma_1) + (\delta m|\sigma_2),
\label{Eq:deltam}
\ee
where $\log \bar{f}={\sum_i \log f_i / N}$, $i=1, ..., N$ and $N$ is
the total number of pixels in the magnification map. By this definition, 
$\delta m_i$ reflects mainly
the effect of microlensing. The light curve of a system at short
wavelength (e.g., $0.16\um$, or $0.63\um$) is more significantly
affected by the microlensing event than that at long wavelength (e.g.,
$3.16\um$; see the blue, red and green lines in each panel of
Figure~\ref{fig:f5}).  The maxima in the light curves are resulted
from caustic crossing events. The heights of these maxima are
sensitive to the ratio of the caustic size to the size of the emitting
region, i.e., the larger the value of this ratio, the larger the
height. The width of the caustic crossing event increases with
increasing wavelength $\lambda$ and decreasing effective relative
velocity $v_{\rm e}$.

Figure~\ref{fig:f5} also shows the difference between the light curves
obtained for the BBH systems ($\delta m^{\rm BBH}_i$) and those
obtained for the corresponding single disk systems ($\delta m^{\rm
S}_i$), i.e., 
\be
\Delta \delta m_i=\delta m^{\rm BBH}_i - \delta m^{\rm S}_i,
\ee
where $\delta m^{\rm BBH}_i$ is the light curve obtained for a BBH
system, $\delta m^{\rm S}_i$ is the light curve obtained for a
corresponding single disk system, which has a single central MBH with
mass $\mbh =\mbhp+\mbhs$ and accretion rate $f_{\rm E} = f_{\rm E,c}$.
For the system S1, $\Delta \delta m_i=\delta m^{\rm S0}_i -
\delta m^{\rm S}_i$, where $\delta m^{\rm S0}_i$ and $\delta
m^{\rm S}_i$ are generated by different random numbers, therefore,
$\Delta \delta m_i$ for the system S0 only represents the contribution
from the Gaussian intrinsic variations of the lensed QSO and the
measurement errors. As seen from
Figure~\ref{fig:f5}, the difference is significant during the caustic
crossing event at $\lambda=0.16\mu$m and $0.63\mu$m for most of the
BBH systems except the BBH system B3 and B4. The light curves for the
BBH system B3 and B4 do not significantly deviate from those of the
corresponding single disk systems because the emission region for
photons at $\lambda=0.16\mu$m and $0.63\mu$m are well within the
gap. For all BBH systems, the differences are not significant at
$\lambda=3.16\mu$m, because the emission region for most photons at 
this wavelength is well outside of the gap and the size of this region
is substantially larger than the caustic size.  

For a BBH system, if the brightness of the secondary (or primary) disk
is negligible compared to that of the primary (or secondary) disk,
then the gap may have a significant effect on limiting the width of
the spike in the light curve caused by the caustic crossing event. For
the system S0, for example, the width of the spike in the light curve
at $\lambda= 0.16\um$ is narrower than that at $\lambda= 0.63 \um$
(see Figure~\ref{fig:f5}); while the width of the spike at $\lambda
=0.16\um$ is almost the same as that at $\lambda = 0.63\um$ for the
system B1. The main reason for the latter case is that the gap in the
disk limits the size of the emitting region of $0.63\um$ photons. 

If the brightness of the secondary disk is comparable to that of the
primary disk, the light curves resulted from the BBH system could be
more complicated.  In the system B7, the two disks associated with the
two BBH components have comparable brightness. For some trajectories,
each lens may pass through two sources one after the other, and the
resulting light curves may thus have two or more spikes. To illustrate
these special cases, we assign three more trajectories for the system
B7 (white lines from left to right in the panel B7 of
Figure~\ref{fig:f4}, denoted as trajectory (b), (c), and (d),
respectively), in addition to the trajectory assigned for all the
systems. Correspondingly, we assign three more trajectories in
the case of the single disk system S0 (white lines from left to right
in the panel S0 of Figure~\ref{fig:f4}, also denoted as trajectory
(b), (c), and (d), respectively). As shown in Figure~\ref{fig:f6}, the
mock light curves obtained from these trajectories can have more
complicated structure, e.g., two or three spikes, as expected.

\subsection{The case of rotating surface brightness distributions}
\label{subsec:lcrotate}

In a more general case, the surface brightness distribution of a BBH
system changes with time particularly at the region within the gap
because of the rotation of the BBH and its associated disks around the
center of mass. In this section, we assume $v_{\rm e} \simeq 10^3
\kms$, and $\Delta t= t_{\rm E}= \re /v_{\rm e}\simeq 10~{\rm yr} \gg
t_{\rm orb}$ for $\mbh \simeq 10^8\msun$ and $\abbh \sim 1000\rg$, and
thus the light curves of these BBH systems can be significantly
affected by the BBH motions (see discussion in
section~\ref{subsec:nonrotating}).  At each monitoring time, the
surface brightness distribution of the BBH source is different from
that at a previous time. Therefore, we do not have a fixed convolved
magnification map for a system. We assign trajectories to the original
magnification map for each system listed in Table~\ref{tab:t1},
similar to that shown in Figure~\ref{fig:f4}, and then perform the
convolution of the original magnification map with the surface
brightness distribution map at each monitoring time to obtain the
light curves for each system. We also add the intrinsic variations and
the measurement errors to the mock light curves in the same way as
that described in section~\ref{subsec:nonrotating}. 

Figures~\ref{fig:f7} and \ref{fig:f8} show the obtained light curves
by considering the BBH rotation, which correspond to those shown
in Figure~\ref{fig:f5} and \ref{fig:f6} for the case of static surface
brightness distributions. Comparing with the light curves shown in
Figure~\ref{fig:f5} (or \ref{fig:f6}), those light curves shown in
Figure~\ref{fig:f7} (or \ref{fig:f8}) have much more complicated
structure because of the BBH rotation.  For example, there are six
peaks in the light curves at $\lambda= 0.16\mu$m and $0.63\mu$m for
the system B7, which is mainly caused by the BBH rotation.  During the
caustic crossing event, the disk1 (or disk2) first crosses the caustic
and moves away from it, following that the disk2 (or disk1) crosses
the caustic after a time roughly half of the BBH period, and then the
disk1 comes back to cross the caustic again at a time roughly of one
BBH period.  These quasi-periodic caustic crossing events of the
disk1 and disk2 finish until the caustic crosses the whole gap region,
which may consequently result in many peaks in the light curves as the
emission from disk1 and disk2 are comparable. The number of the peaks
depends on the ratio of the BBH orbital period to the time needed for
the caustic crossing the whole gap region. (For similar rotation effects
in Galactic microlensing, see \citealt{Penny11}.)  For the system B8, there
are only four peaks in the light curves (at $\lambda=0.16\mu$m and
$0.63\mu$m), which is different from that for the system B7. The
reason is that the emission from the disk1 is negligible (even
when amplified by microlensing) and the disk2 crosses the caustic again
roughly one BBH period after a previous crossing.  Similarly, the
structures of the light curves for other systems shown in
Figures~\ref{fig:f7} and \ref{fig:f8} can also be understood.

In summary, we conclude that the rich structures in the microlensing
light curves of lensed BBH QSOs shown in Figures~\ref{fig:f5}-
\ref{fig:f8} may help to probe and reveal the nature of BBH systems in
lensed QSOs. For BBH systems that show strong rotation effects, even a
single-color microlensing light curve is already highly suggestive of
their existence.

\section{Disk half light radius--wavelength relation estimated from
the mock light curves} \label{sec:disksize}

\subsection{Fitting Method}\label{subsec:method}

The size of the emitting region of the photons at a given wavelength
can be extracted from the mock light curves of each lensed QSO
generated in section~\ref{sec:mocklc} \citep[see][]{Kochanek04}. In
reality, it is not clear at all which observed lensed QSO is due to a
BBH system.  Therefore, we first adopt the single disk model described
in section~\ref{subsec:sd} to fit the mock light curves, although most
of them are generated by assuming a background BBH source. We then
check whether the $r_{1/2}-\lambda$ relationship estimated from the
light curves of a BBH system is significantly different from that
expected from a single MBH system, and whether the difference, if any,
can be used to select BBH candidates from the lensed QSOs.

According to the standard thin accretion disk model, the surface
brightness profile at any given narrow band with a central wavelength
$\lambda$ can be described by 
\be 
I(r) \propto \left[{\rm exp}\left((r/r_{\rm s})^{3/4}
(1-\sqrt{r_{\rm in}/r})^{1/4}\right)-1\right]^{-1},
\label{eq:bright}
\ee
where $r$ is the distance to the central MBH,  $r_{\rm s} = \left(
\frac{\lambda k_{\rm B}}{h c}\right)^{4/3}\left(\frac{3G^2 m_{\rm H}
\mbh^2 f_{\rm E}}{2\epsilon c\sigma_{\rm T}\sigma_{\rm B}
}\right)^{1/3} \approx 0.41 r_{1/2}$, and $r_{1/2}$ is the half light
radius (see Equation~\ref{eq:rhalf}). Photons in the optical to near
infrared bands are mainly emitted from the disk region with $r \gg
r_{\rm in}$, for which the term $(1- \sqrt{r_{\rm in}/ r})^{1/4}$ in
Equation~(\ref{eq:bright}) can be approximated as a constant. Therefore,
the brightness distribution at a given wavelength (or a narrow band
filter) of the disk is roughly determined by the single parameter
$r_{\rm s}$ or equivalently $r_{1/2}$. We choose $30$ different
brightness profiles for each source, and these profiles are described
by Equation~(\ref{eq:bright}) by setting $r_{1/2}$ to be in the range
of $10$ to $10^{-3}\re$ with equal logarithmic intervals.

\begin{figure*}
\centering
\includegraphics[scale=0.73,width=6.6cm,angle=-90]{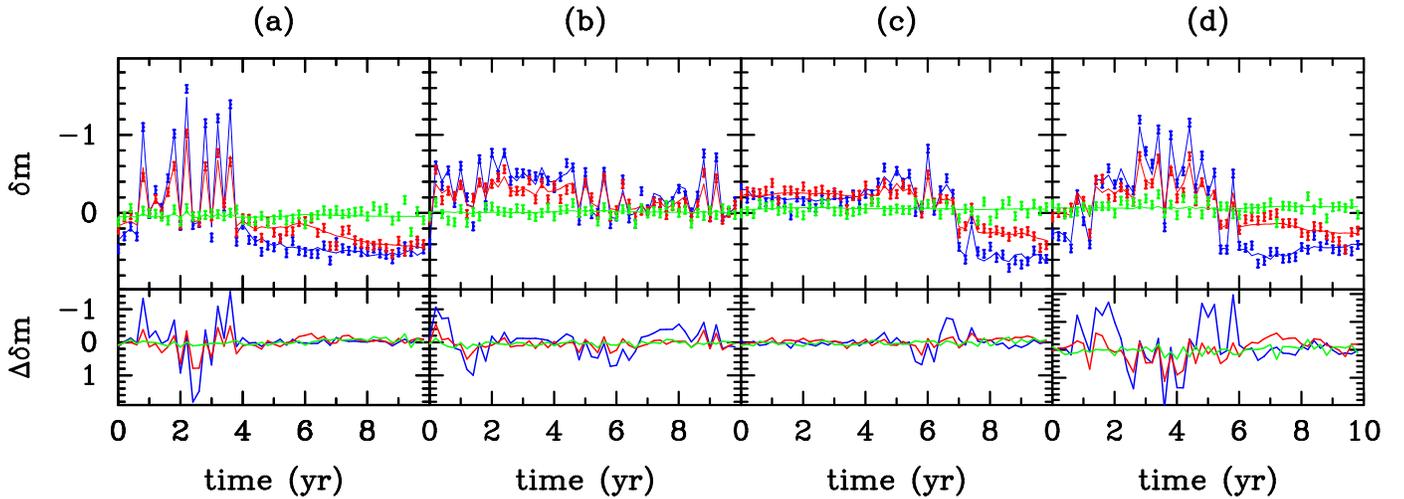}

\caption{Mock light curves obtained for the BBH system B7 by
considering the BBH rotation. Legends are similar to those for the
bottom panels of Figure~\ref{fig:f6}.  }

\label{fig:f8}
\end{figure*}

For the lens, we set the convergence $\kappa$ and shear $\gamma$ to be
the same as those used to obtain the mock light curves in section
\ref{sec:mocklc}. The reason is that $\kappa$ and $\gamma$ can be well
constrained by observations on the macro-lensing of the background
QSO. The contribution of stars to the convergence $\kappa_*$ may also
be constrained by detailed observations on the lens galaxy. However,
it is practically difficult to constrain the effective relative
velocity $v_{\rm e}$ through observations other than the microlensing.
For simplicity, nevertheless, we set $\kappa_*$ and $v_{\rm e}$ the same as
the initial input values to generate the mock light curves in the
following Bayesian fitting to obtain the size of the emitting region,
if not otherwise stated.  In principle, $\kappa_*$ and $v_{\rm e}$ can be
simultaneously obtained from the fitting of the light curves, which
will be discussed in section~\ref{subsec:kappave}.  For general use,
we keep $\kappa_*$ and $v_{\rm e}$ as free parameters in the following
formalization of the Bayesian fitting.  

According to the above settings, we generate four different original
magnification maps for the lens. (1) For the case of static surface
brightness distributions, we obtain $120$ convolved magnification maps
by convolving with the $30$ brightness profiles set above, each of
which is stored in $2048\times 2048$ arrays with a side length of
$40\left<\re \right>$ and a pixel size of $\sim 0.02\left<\re\right>$.
We note that the magnification maps here are generated by using random
numbers different from that set to obtain the mock light curves in
section~\ref{sec:mocklc}. We can then randomly choose trajectories
(both the starting points and directions) in each of the convolved
magnification maps, and the trajectory length is determined by the
effective relative velocity and the ``monitoring'' period of the mock
light curves. (2) For the case of rotating surface brightness
distributions, we randomly choose trajectories in the original
magnification map, and obtain $\delta m'_i$ at each monitoring time by
convolving the original magnification map with the surface brightness
distribution of the source at that time.  Similar to that in section
\ref{sec:mocklc} for obtaining the mock light curves, a model light
curve $\delta m'_i$ can be obtained for each trajectory. We use these
light curves to fit the mock light curves by using the standard
$\chi^2$ statistics, i.e.,
\be
\chi^2= \sum_i { \left( \delta m_i - \delta m'_i \right)^2
              \over \sigma_{i}^2 + \sigma_1^2 },
\label{eqn:chi1}
\ee
where $\sigma_{i}$ is set to be $\sigma_2$. By searching for the
minimum of the $\chi^2$ value, we may then obtain a constraint on the
half light radius at any given wavelength for each system as follows.

Using the Bayesian theorem, the posterior probability distribution of the 
parameters involved in the fitting for a given set of data \{$D$\} is 
\bea
P({\hat{r}_s,\kappa_*,\hat{v}_e},{\mathbf{\xi_t}}|D) &\propto &
P( D | {\hat{r}_s,\kappa_*,\hat{v}_e},{\mathbf{\xi_t}})
P({\hat{r}_s})P({\kappa_*})\times \nonumber \\
& & P({\hat{v}_e})P({\mathbf{\xi_t}}),
\label{eq:postprob}
\eea
where $P(\hat{r}_s)$, $P(\kappa_*)$, $P(\hat{v}_e)$, and $P({\mathbf{\xi_t}})$
describe the prior probability estimates for the physical parameters 
($\hat{r}_s$,  $\kappa_*$, $\hat{v}_{\rm e}$), and the trajectory variables
(${\mathbf{\xi_t}}$), respectively, and 
\be
P(D|{\hat{r}_s,\kappa_*,\hat{v}_e},{\mathbf{\xi_t}}) =P(\chi^2)\propto 
\Gamma\left[ { N_{\rm dof}-2 \over 2 }, { \chi^2 \over 2f_0^2 } 
\right],
\ee
$\Gamma[a,b]$ is an incomplete Gamma function, $N_{\rm dof}$ is the
degree of freedom, and $f_0$, in the range from 0 to 1, is a scale
parameter to control the magnitude uncertainties of the light curves
(see Equation 9 in \citealt{Kochanek04}). All the Bayesian parameter
estimates are normalized by the requirement that the total probability
is unity, i.e.,
\be
\int d{\hat{r}_s}d{\kappa_*}d{\hat{v}_e}d{\mathbf{
\xi_t}} P({\hat{r}_s,\kappa_*,\hat{v}_e},{\mathbf{\xi_t}}| D)=1.
\ee 
The starting points and directions of trajectories,
${\mathbf{\xi_t}}$, are randomly chosen in the convolved (or
original) magnification map, which are nuisance variables. The
posterior joint probability distribution of the physical parameters
(${\hat{r}_s}$, $\kappa_*$, $\hat{v}_{\rm e}$) can be obtained by
integrating Equation~(\ref{eq:postprob}) over ${\mathbf{\xi_t}}$,
i.e.,
\be
P({\hat{r}_s,\kappa_*,\hat{v}_{\rm e}}|D) \propto \int
P({\hat{r}_s,\kappa_*,\hat{v}_{\rm e}},{\bf \xi_t}|D) d{\mathbf{\xi_t}}.
\label{eq:probint}
\ee
We adopt logarithmic priors for the the size parameter $r_{s}$
involved in Equation~(\ref{eq:postprob}), i.e., $P(\hat{r}_s) \propto
1/\hat{r}_s$. 

\begin{figure}
\centering
\includegraphics[scale=1.0,width=10.0cm,angle=-90]{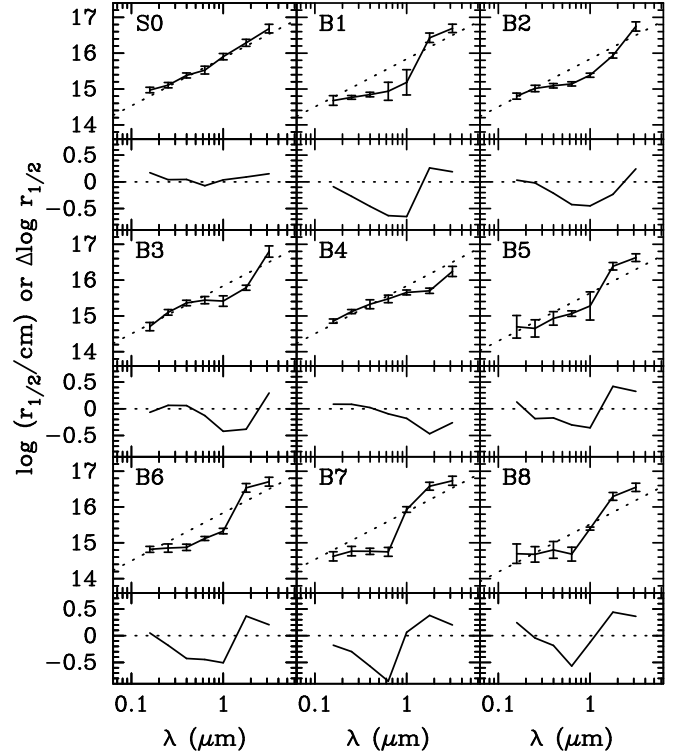}

\caption{Half-light radius-wavelength ($r_{1/2}-\lambda$) relation
estimated from the mock light curves for each BBH system (and the
single MBH system S0) shown in Figure~\ref{fig:f5} for static surface
brightness distributions. In those panels with a label, the points
with error bars and the curve connecting them show the
$r_{1/2}-\lambda$ relation obtained for the system S0, B1, B2, B3, B4,
B5, B6, B7 and B8, respectively (see Table~\ref{tab:t1}); the
dotted line represents the expected $r_{1/2}-\lambda$ relation,
according to Equation~(\ref{eq:rhalf}), for a corresponding single MBH
system with a mass the same as the total mass of the BBH system and
accretion rate the same as that of the circum-binary disk. In each
small panel, right below each big panel with a label, the curve
represents the difference between the obtained $r_{1/2}-\lambda$
relation and the ones expected for the corresponding single MBH
system, which are shown in each big panel.  }

\label{fig:f9}
\end{figure}

To speed up the fitting process, we also adopt a method similar to
that in \citet{Kochanek04}: we first try to find $10^3$ trial
trajectories, of which the corresponding model light curves satisfy
$\chi^2\leq \chi^2_{\rm max}=3N_{\rm dof}$; and then locally
optimize the starting points and directions of trajectories in the
vicinity of each trial trajectory to search for the minimum value of
$\chi^2$. Mock light curves at short wavelength are more complicated
than those at long wavelength, many more test trajectories are needed
to find $10^3$ trial trajectories for the mock light curves at short
wavelength than for that long wavelength.  We adopt $\sigma_1$ the
same as the input ones initially if $10^3$ trial trajectories with
$\chi^2\leq 3 N_{\rm dof}$ can be found among the search of
no more than $10^9$ trajectories and a good fit can be obtained,
otherwise, we increase $\sigma_1$ until $10^3$ trial trajectories can
be found among $10^9$ trajectories. 

It has been assumed that four images (A, B, C, and D) are generated by
the macrolensing for each system, and correspondingly four light
curves, independent from each another, are generated for these
images.\footnote{The shears of the four images may be correlated with
each other, which should be taken into account when fitting light
curves of different images of a system. In this paper, we ignore the
correlation among the shears for the four images of a system, for
simplicity.} Therefore, we combine the results obtained from each
light curve together by multiplying the probability, i.e.,
\be
P_{\rm tot}(\hat{r}_s,\kappa_*,\hat{v}_{\rm e}|D)=\prod\limits_j 
P_{j}(\hat{r}_s,\kappa_*,\hat{v}_{\rm e}|D),
\ee 
where $P_j$ is obtained from Equation~(\ref{eq:probint}) for a light curve
$j$, and $j=1$, $2$, $3$, and $4$, representing the light curve obtained 
for the image A, B, C and D, respectively.

\begin{figure}
\centering
\includegraphics[scale=1.0,width=5.25cm,angle=-90]{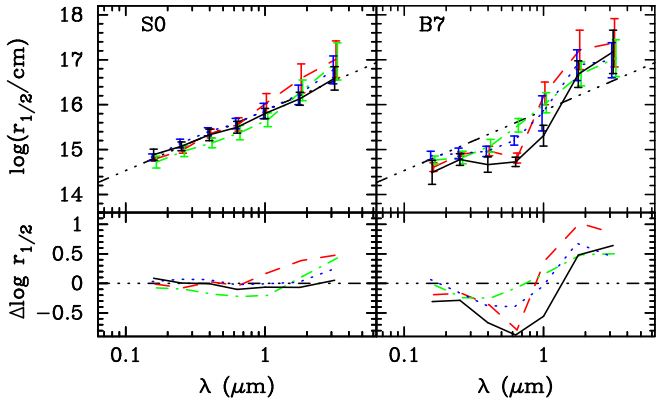}

\caption{Half-light radius-wavelength relation estimated from the mock
light curves of the image A shown in Figure~\ref{fig:f6}. In each
panel, black solid, red dashed, green dot-dashed and blue dotted lines
represent the results obtained for light curves correspondingly shown
in Figure~\ref{fig:f6} for trajectory (a), (b), (c) and (d),
respectively. In the top panels, the dot-dot-dot dashed lines
represent the half-light radius-wavelength relation obtained for the
corresponding single MBH QSO systems. }

\label{fig:f10}
\end{figure}

The posterior probability distributions for individual physical
parameters can be obtained by marginalizing $P_{\rm tot}$ over
remaining parameters, for example, the posterior probability
distribution of the source size is
\begin{equation}
P({\hat{r}_s}|D) \propto \int P_{\rm tot}({\hat{r}}_s,\kappa_*,
\hat{v}_{\rm e}|D) d\kappa_* d\hat{v}_{\rm e}.
\end{equation}
By adopting a Gaussian profile to fit $P({\hat{r}_s}|D)$, we obtain
the most probable size of the emitting region and its uncertainty as
the peak location of the Gaussian profile and its dispersion. Similar
to the source size, the posterior probability distribution of the
stellar convergence $\kappa_*$ (or the effective relative velocity
$\hat{v}_{\rm e}$) may also be obtained by marginalizing $P_{\rm tot}$
over other remaining parameters (see discussions in
section~\ref{subsec:kappave}).

\subsection{Fitting Results: half-light radius-wavelength relation}
\label{subsec:results}

\subsubsection{The case of static surface brightness distributions}
\label{subsubsec:static}

Figure~\ref{fig:f9} shows the $r_{1/2}-\lambda$ relations obtained
from the mock light curves shown in Figure~\ref{fig:f5} through the
Bayesian fitting (symbols with errorbars) by assuming a single disk
model for all the systems listed in Table~\ref{tab:t1}. As seen from
the top left panel of Figure~\ref{fig:f9}, the $r_{1/2}-\lambda$
relation obtained for the single disk system S0 is well consistent
with the simple expectation from Equation~(\ref{eq:rhalf}) (solid line),
which justifies the Bayesian fitting method presented in
section~\ref{subsec:method}. In each panel for a BBH system in
Figure~\ref{fig:f9}, the dotted line represents the expected
$r_{1/2}-\lambda$ relation for a corresponding single MBH system, in
which the central MBH mass and accretion rate is the same as the total
MBH mass and total accretion rate of the BBH system, respectively. For
the BBH systems, the estimated $r_{1/2}-\lambda$ relations deviate
from that expected from the corresponding single MBH systems, i.e.,
$r_{1/2}\propto \lambda^{4/3}$.  The main difference is that the size
of the emitting region changes with increasing wavelength much slower
than $r_{1/2} \propto \lambda^{4/3}$ (or almost does not change) in
some wavelength ranges (see the panels for system B1, B5, B6, B7, B8).
For B1, B6, B7, and B8, $r_{1/2}$ is almost a constant in the range of
$\lambda\sim 0.16 - 0.63\um$ (see Figure~\ref{fig:f9}), while it is
expected to change by a factor of $6$ in the case of the single
standard thin disk (see Equation~\ref{eq:rhalf}). This difference is
mainly due to the existence of a gap in the BBH systems, which limits
the emitting area of photons with wavelength in a certain range. From
B1, B2, B3, to B4, this feature moves toward longer wavelengths and
becomes less prominent mainly because the gap moves to larger radii.
At longer wavelength ($\ga 1\um$), the half-light radii for each BBH
system jump up to be even larger than that expected for the
corresponding single disk system, which is due to the main emitting
region jump from the inner disk(s) to the outer circum-binary disk
and the lack of emission in the gap region.

We also show the residuals for the $r_{1/2}-\lambda$ relation, i.e.,
the difference between the data points and the dotted line in each
panel with a label, in a small panel below each labeled panel in
Figure~\ref{fig:f9}, respectively.  As seen from the residual curves,
there is a dip in each curve for a BBH system because the gap opened
by the secondary MBH limits the emitting area of the photons in a
certain wavelength range. Out of this wavelength range, the effect of
the gap is negligible and the $r_{1/2}-\lambda$ relation of a BBH
system is similar to that for a single MBH system.  With increasing
separation of the two MBHs, the dip moves toward longer wavelength
(see Figure~\ref{fig:f2} for the system B1, B2, B3 and B4) because the
gap moves outward and the short wavelength photon emitting region is
correspondingly less affected by the gap. According to the panels B5,
B6, B7 and B8 in Figure~\ref{fig:f9}, the feature of a dip in the
residual curve is maintained if choosing somewhat different mass ratio
$q$ and Eddington ratios $f_{\rm E,1}$ and $f_{\rm E,2}$. These dips
are unique features of those BBH systems with physical parameters at
some certain ranges and  can be taken as indicators of the existence
of BBHs in the background lensed QSOs.

For some light curves, e.g., B7 (b) in Figure~\ref{fig:f6}, B7 (a) in
Figure~\ref{fig:f8}, a good fit cannot be obtained if we set
$\sigma_1$ the same as the input ones to generate the mock light
curves. To solve this problem, we increase $\sigma_1$ until a good fit
is obtained and then take the resulting $r_{1/2}$ as the best fit (see
section~\ref{subsec:method}).  Figure~\ref{fig:f10} also shows the
$r_{1/2}-\lambda$ relation obtained for the light curves only for the
image A shown in Figure~\ref{fig:f6} (for trajectories (b), (c)
and (d) in the panels S0 and B7 of Figure~\ref{fig:f4}). For example,
$\sigma_1$ is set to be $0.05$ for the light curves at short
wavelengths in panels (a) and (c) in Figure~\ref{fig:f6} for the
system B7, while it is set to be $0.1$ for the light curves at short
wavelength in panels (b) and (d) in Figure~\ref{fig:f6} for the same
system. As seen from Figure~\ref{fig:f10}, there are also clear dips
in the residuals for the $r_{1/2}-\lambda$ relation estimated from the
light curves, corresponding to trajectories (b), (c) and (d), for the
BBH system B7, similar to those shown in Figure~\ref{fig:f9}. These
dips can also be taken as indicators of the existence of a BBH in the
background lensed QSO system.

\subsubsection{The case of rotating surface brightness distributions}

\begin{figure}
\centering
\includegraphics[scale=1.0,width=10.0cm,angle=-90]{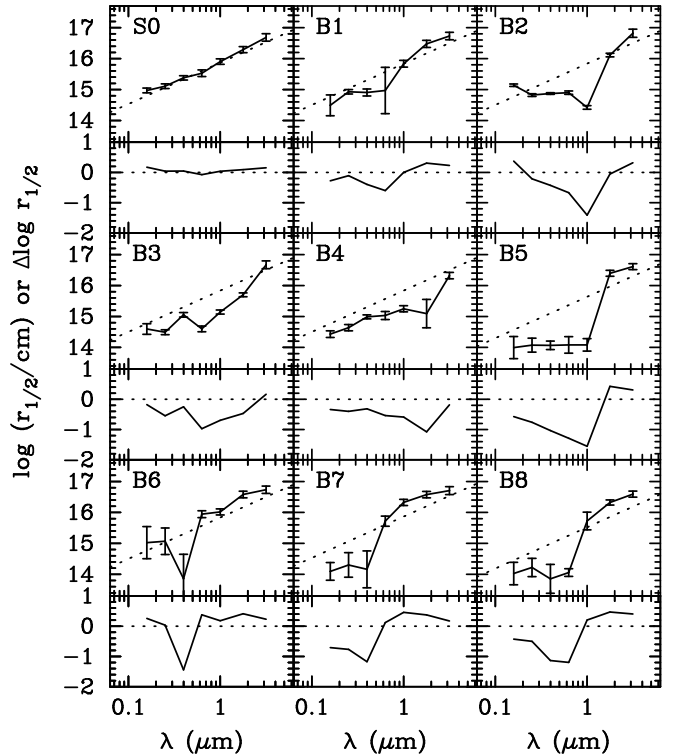}

\caption{Half-light radius-wavelength relation estimated from the mock
light curves for each BBH system (and the single MBH system S0) shown
in Figure~\ref{fig:f7}. Legends are similar to those for
Figure~\ref{fig:f9}. }

\label{fig:f11}
\end{figure}

\begin{figure}
\centering
\includegraphics[scale=1.0,width=5.25cm,angle=-90]{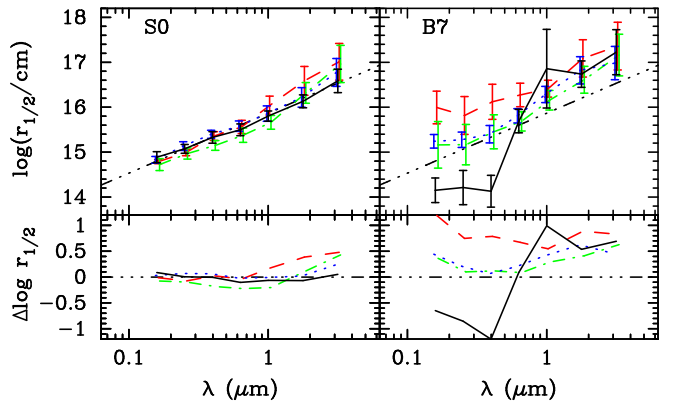}

\caption{Half-light radius-wavelength relation estimated from the mock
light curves for the single MBH system S0 and the BBH system B7 shown
in
Figure~\ref{fig:f6}. Legends are similar to those for
Figure~\ref{fig:f10}.
}

\label{fig:f12}
\end{figure}

We adopt the same method as that for the case of static surface
brightness distributions in section~\ref{subsubsec:static} to fit the
light curves obtained for the case of rotating surface brightness
distributions (see Figures~\ref{fig:f7} and \ref{fig:f8}).
Figure~\ref{fig:f11} shows the $r_{1/2}-\lambda$ relations estimated
from the light curves shown in Figure~\ref{fig:f7}, of which the
rotation of BBHs and their associated disks are considered. These
estimated $r_{1/2}-\lambda$ relations look very similar to those shown
in Figure~\ref{fig:f9} for the case of static surface brightness
distributions and the main feature of a dip is still present in each
residual curve for BBH systems, which suggest that this feature can be
always taken as an indicator of the existence of a BBH in a lensed
QSO. Figure~\ref{fig:f12} shows the $r_{1/2}-\lambda$ relations
obtained from the light curves only for the image A shown in
Figure~\ref{fig:f8}. Some light curves, such as those for trajectories
(b), (c) and (d) shown in Figure~\ref{fig:f8}, may rapidly change
during the caustic crossing event because of the rapid rotation of the
surface brightness distributions within the gap and cannot be well
fitted if we assume $\sigma_1$ is the same as the input values to
generate these curves.  Increasing $\sigma_1$ in the fitting is
equivalent to smooth the light curves, which results in bigger
half-light radii. As seen from Figure~\ref{fig:f12}, the estimated
$r_{1/2}$ for the light curves corresponding to trajectories (b), (c)
and (d) are larger than that expected from the corresponding single
disk systems by a factor of $\sim 2-4$. These cases might be
responsible for some of the sources that have flatter
$r_{1/2}-\lambda$ relations and relatively larger disk sizes comparing
with the expectations from the standard thin disk model, as discovered
by \citet{Blackburne11}. 

\subsubsection{Exploration of the parameter space}

We further explore the parameter space for BBH systems by varying the
values of $a_{\rm BBH}$, $M_{\bullet}$, $q$, $f_{\rm E,1}$, $f_{\rm
E,2}$. As demonstrated above, the basic feature of the  estimated
$r_{1/2}-\lambda$ relation for BBH systems by considering the BBH
rotation is similar to that without considering BBH rotation.  For
simplicity, therefore, we only consider the case of static surface
brightness distributions in this sub-section. The trajectories adopted
in the convolved magnification maps are the same as the trajectory (a)
shown in Figure~\ref{fig:f4}.

\begin{figure}
\centering
\includegraphics[scale=1.0,width=8.7cm,angle=-90]{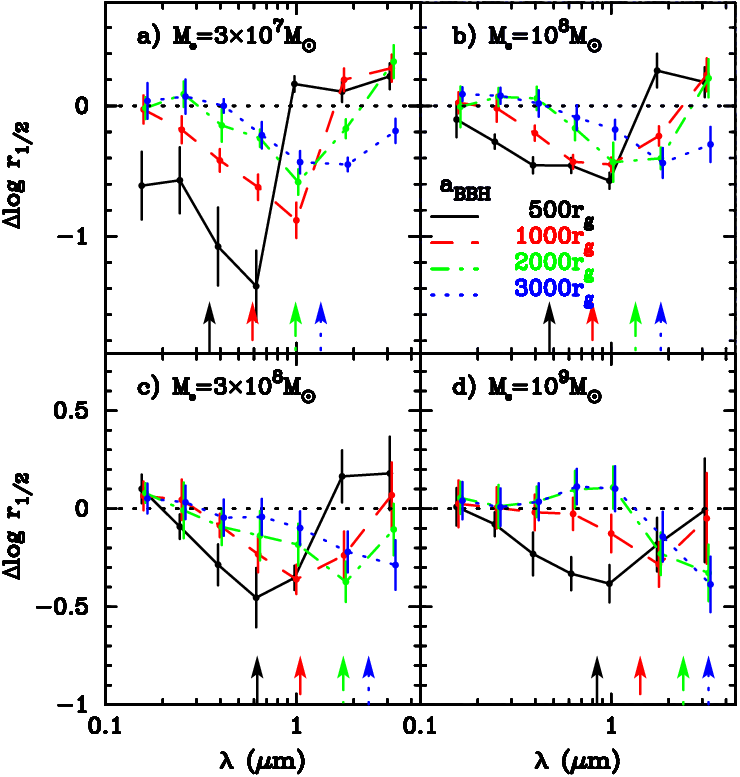}

\caption{Difference between the $r_{1/2}-\lambda$ relation obtained
for BBH systems and that for the corresponding single MBH
systems. Panels (a), (b), (c) and (d) show the BBH system with total
mass $3\times 10^7$, $10^8$, $3\times 10^8$ and $10^9
\msun$, respectively.  The BBH systems shown in each panel have the
same mass ratio ($q=0.25$) and Eddington ratios ($f_{\rm E,1},f_{\rm
E,2})=(0.3, 0.01$) but different separation $a_{\rm BBH}=500\rg$
(black symbols and line), $1000\rg$ (red symbols and dashed line),
$2000\rg$ (green symbols and dot-dashed line), and $3000\rg$ (blue
symbols and dotted line), respectively. Black, red, green and blue
vertical arrows mark the locations of the gap  expected from
Equation~(\ref{eq:rhalf}) if assuming that the half-light radius
$r_{1/2}= \frac{1}{2}\left(\abbh+\frac{1}{2}\left[R_{\rm RL} (1/q) -
R_{\rm RL}(q)\right] \right) -\frac{q a_{\rm BBH}}{ 1+q}$ (i.e., the
distance from the center of the gap to the mass center of the BBH) for
each system.  }

\label{fig:f13}
\end{figure}

Figure~\ref{fig:f13} shows the residuals of the $r_{1/2}-\lambda$
relation for BBH systems with different total mass (i.e., $M_{\bullet}
=3\times 10^7$, $10^8$, $3\times 10^8$, and
$10^9\msun$) at different separations (i.e., $a_{\rm BBH}=500$,
$1000$, $2000$ and $3000\rg$), but the same mass ratio $q=0.25$
and Eddington ratios ($f_{\rm E,1}=0.3$, $f_{\rm E,2}=10^{-4}$).  As
seen from Figure~\ref{fig:f13}, the $r_{1/2}-\lambda$ residual is more
prominent for BBH systems with smaller $M_{\bullet}$ and smaller
$a_{\rm BBH}$, and the wavelength of the dip in the $r_{1/2}-\lambda$
residual increases with increasing $M_{\bullet}$ and increasing
$a_{\rm BBH}$. For BBH systems with $M_{\bullet}=10^8\msun$ and
$a_{\rm BBH}\ga 3000r_{\rm g}$ (or $M_{\bullet}=10^9\msun$ and $a_{\rm
BBH}\ga 500r_{\rm g}$), the dip moves to the infrared band in the QSO
rest frame. Considering that many lensed QSOs are at redshift $z\sim
1-2$, the dip can be observed in the optical bands only for those
systems with small MBH mass ($M_{\bullet}\la 10^8\msun$) and/or small
separations ($a_{\rm BBH}\la 1000r_{\rm g}$).

\begin{figure}
\centering
\includegraphics[scale=1.0,width=8.7cm,angle=-90]{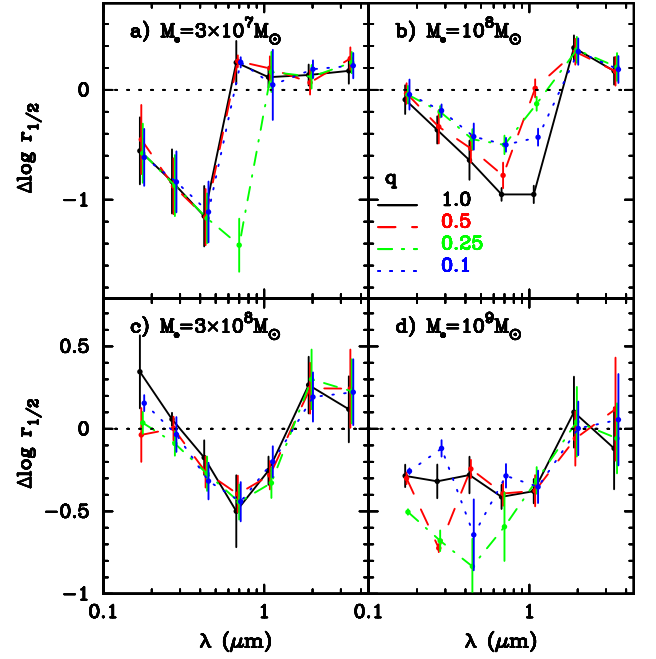}

\caption{Difference between the $r_{1/2}-\lambda$ relation obtained
for BBH systems and that obtained for the corresponding single MBH
systems. Panels (a), (b), (c) and (d) show the BBH systems with total
mass $3\times 10^7$, $10^8$, $3\times 10^8$ and $10^9
\msun$, respectively.  The BBH systems shown in each panel have the
same separation ($a_{\rm BBH}=500\rg$) and Eddington ratios ($f_{\rm
E,1},f_{\rm E,2})=(0.3, 0.3$) but different mass ratio $q=1$ (black
symbols and solid line), $0.5$ (red symbols and dashed line), $0.25$
(green symbols and dot-dashed line) and $0.1$ (blue symbols and dotted
line), respectively. }

\label{fig:f14}
\end{figure}

Figure~\ref{fig:f14} shows the $r_{1/2}-\lambda$ residuals for BBH
systems with different total mass (i.e., $M_{\bullet}= 3 \times
10^7$, $10^8$, $3\times 10^8$, and $10^9\msun$) and
different mass ratio $q$, but the same separation $a_{\rm
BBH}=500r_{\rm g}$ and Eddington ratio $f_{\rm E,1}=0.3$, $f_{\rm
E,2}=0.3$.  As seen from Figure~\ref{fig:f14}, the dips in the
$r_{1/2}-\lambda$ residual curves remain more or less at the same
location for different choices of the mass ratio but the width of the
dip shrinks slightly with decreasing $q$. 

\begin{figure}
\centering
\includegraphics[scale=1.0,width=4.85cm,angle=-90]{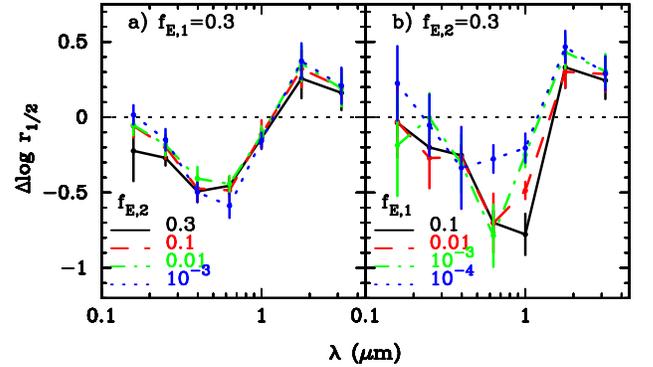}

\caption{Difference between the $r_{1/2}-\lambda$ relation obtained
for BBH systems and that obtained for the corresponding single MBH
systems.  The BBH systems shown here have the same total mass
($M_{\bullet}=10^8\msun$), mass ratio ($q=0.25$), separation ($a_{\rm
BBH}=500\rg$), but different Eddington ratios. In panel (a), black,
red, green and blue symbols (solid, dashed, dot-dashed and dotted
curves) represent ($f_{\rm E,1},f_{\rm E,2})=(0.3, 0.3)$, $(0.3,
0.1)$), ($0.3, 0.01$) and ($0.3, 0.001$), respectively; while they
represent $(0.1, 0.3)$ $(0.01, 0.3)$, ($0.001, 0.3$), ($10^{-4}, 0.3$)
in panel (b), respectively. }

\label{fig:f15}
\end{figure}

Figure~\ref{fig:f15} shows the $r_{1/2}-\lambda$ residuals for BBH
systems with different Eddington ratios ($f_{\rm E,1}$, $f_{\rm E.2}$)
but the same total mass $M_{\bullet}=10^8\msun$, mass ratio $q=0.25$,
separation $a_{\rm BBH}=500r_{\rm g}$. In panel (a), the black points
(solid curve), red points (dashed curve), blue points (dotted curve)
and green points (dot-dashed curve) represent the case with $(f_{\rm
E,1},f_{\rm E,2})=(0.3,0.3)$, $(0.3,0.1)$, $(0.3,0.01)$ and
$(0.3,0.001)$, respectively; while they represent
$(f_{\rm E,1},f_{\rm E,2}=(0.1,0.3)$, $(0.01,0.3)$, $(0.001,0.3)$ and
$(10^{-4},0.3)$ in panel (b), respectively. As seen from the panel (a)
in Figure~\ref{fig:f15}, the location of the dip is insensitive to the
Eddington ratio of the secondary disk as the emission is dominated by
the primary disk. The dip may move toward slightly smaller wavelength
if the emission from the secondary disk becomes dominant as shown in
the panel (b). The reason is that the mass of the secondary MBH is
smaller than the primary ones (see Equation~\ref{eq:rhalf}).

\subsection{Stellar Convergence and Effective Relative Velocity }
\label{subsec:kappave}

In the above Bayesian fitting processes, we adopt fixed $\kappa_*$ and
$v_{\rm e}$. In principle, these two parameters can also be
simultaneously fitted, which may introduce some degeneracy in the
fitting parameters. To test this, we choose four different values for
the stellar convergence (i.e., $\kappa_*=0.125\kappa, 0.25\kappa,
0.5\kappa, \kappa$) and five different values for the effective
relative velocity (i.e., $\hat{v}_{\rm e} = 10^{-1.5}, 10^{-1.25},
10^{-1}, 10^{-0.75}, 10^{-0.5}\re/{\rm yr}$), to simultaneously obtain
estimates of $\kappa_*$ and $\hat{v}_{\rm e}$ by using the mock light
curves. 

According to the best fits to the mock light curves of the single MBH
system in section~\ref{subsec:results}, we find that the best-fit
values of the stellar convergence $\kappa_*$ and the effective
relative velocity $v_{\rm e}$ are well consistent with the input ones.
For the BBH systems, we find that majority of the best fits,
especially those cases at short wavelength, can recover the input
$\kappa_*$ and $v_{\rm e}$ well; while some of the best fits for those
cases (especially for the system B7) at long wavelength may give a
$\kappa_*$ or $v_{\rm e}$ larger than the input ones. At short
wavelength, the stellar convergence and effective relative velocity
can be well recovered from the mock light curves mainly because sharp
structures in the mock light curves, due to caustic crossing events,
encode sufficient information of both quantities; at long wavelength,
however, the mock light curves are flat because of relatively less
significant microlensing effect, and thus offer less capability of
putting accurate constraints on the two quantities (see
Figure~\ref{fig:f5}). However, we find that the constraints on
the disk size and especially the behavior of the $r_{1/2}-\lambda$
relations  are not significantly affected by the uncertainties in the
constraints of $\kappa_*$ and $v_{\rm e}$.

In the system B7, the disk associated with the secondary MBH has a
luminosity comparable to that associated with the primary MBH. For
some trajectories in the convolved magnification map (see
Figure~\ref{fig:f4}), a star may pass through the two disks one after
the other, which leads to two peaks in the mock light curves. Similar
light curves, with two peaks over a similar observational time period,
can also be produced by a background source composed of a single disk
but a lens with higher $\kappa_*$ and/or higher $v_{\rm e}$ as two
stars may pass through the background single disk in the observational
period. As discussed in section~\ref{subsec:results}, the amplitude of the
fluctuations in the light curve is mainly determined by the source
size but not $\kappa_*$ and $v_{\rm e}$. The estimates of the source
size in section~\ref{subsec:results} are not affected much by the degeneracy
existed in the fitting. We note here that the degeneracy described
above may be broken up if more than one caustic crossing events are
observed, as the two peaks due to the two disks are always associated
with each other and come together but the two peaks due to large
$\kappa_*$ and $v_{\rm e}$ are independent. 

\section{Other complications and discussions}
\label{sec:complications}

For demonstration purpose, a number of simplifications for the BBH
systems are made in the above calculations. Below we will discuss the
possible effects of these simplifications on the estimated
$r_{1/2}-\lambda$ relation and probing the signals of BBH systems.

\begin{itemize}

\item Eccentric BBH orbit. We have assumed that the BBH systems
are on circular orbits. However, the eccentricity of BBH systems may
be excited/de-excited to a moderate value during the orbital decay of
BBHs in gaseous disks according to some numerical simulations
\citep[e.g.,][]{MM08, Hayasaki07,Roedig11}. If the orbits of sub-parsec
BBH systems are really significantly eccentric, the gap opened by the
secondary MBH may be larger (or smaller) than that in the
corresponding case with a circular BBH orbit if the BBH orbit is
prograde (or retrograde)  \citep[e.g.,][]{Roedig13}.  Comparing with
that of a circular BBH, the disk associated with each component in the
case of an eccentric BBH should be smaller and the inner edge of the
circum-binary disk should be larger, and thus the anomalies in the
$r_{1/2}-\lambda$ relations of the BBH systems become more (or less)
prominent.

\item Periodical variations of the accretion rates due to the BBH
dynamics. In the above calculations, the light emitted from each BBH
system was assumed to not vary significantly. However, as suggested by a
number of recent studies \citep[e.g.,][]{MM08, Haiman09, Sesana12,
Hayasaki12, DOrazio12}, the accretion onto active eccentric BBH
systems may vary on a timescale on the order of the orbital periods
due to the BBH dynamics. This variation is periodic and may be
substantial, and is totally different from the intraday intrinsic
variation assumed in section~\ref{sec:mocklc}. The variation due to
the BBH dynamics could be mixed with the variation due to microlensing
effect. The rotation of BBHs and their associated disks within the gap
may also cause quasi-periodic variation during the caustic crossing
events but this quasi-periodic variation finishes after the caustic
crosses the whole gap region as discussed in
section~\ref{subsec:lcrotate}. If the period of a BBH system is
sufficiently long compared with the caustic crossing effect, then
there is no much effect caused by the BBH dynamics on the microlensing
light curve. If the period of the BBH systems and the time for the
caustic to cross the whole gap region is relatively short compared to
the light curve period, then variations due to the BBH
dynamics and the microlensing effect can be separated by modeling the
light curves.

\item Non-coplanar disks associated with BBH systems. It is possible
that the BBH orbital plane is mis-aligned with the disk plane.  In a
mis-aligned BBH system, it may be difficult to open a  gap in the
primary disk by the secondary MBH \citep[see][]{Hayasaki12},
therefore, the difference between the $r_{1/2}-\lambda$ relation of a
mis-aligned BBH system and that of a single disk system may thus be
not significant. However, the exact structure of disks associated with
BBH systems is still not fully understood. Future high resolution
simulations, following the dynamical evolution of gas and BBHs from
the galactic to disk scale, may be able to answer whether sub-parsec
active BBH systems have coplanar or non-coplanar disk structures.

\item Disk model. A standard thin disk model is adopted in the above
calculations. In reality, thin disk model may be too simple even for
the single disk case. In order to check the effect on the resulting
$r_{1/2}-\lambda$ relation by the usage of different disk models, we
also adopt the TLUSTY model to estimate the $r_{1/2}-\lambda$ relation
for the systems listed in Table~\ref{tab:t1}. In the TLUSTY model, the
relativistic effects, the disk vertical structure and the radiative
transfer in the disk are simultaneously considered \citep[for details,
see][]{Hubeny00}.  We find that the estimated size at a given optical
wavelength by adopting the TLUSTY model is slightly larger than that
obtained by adopting the standard thin disk model, but the general
shapes of the $r_{1/2}-\lambda$ relations estimated by adopting the
two different disk models are the same. If the area of emitting region
increases rapidly with increasing wavelength in any adopted disk
model, the anomaly in the $r_{1/2}-\lambda$ relation of BBH systems
always exists as a reflection of the gap, a unique and
generic property of accretion flows onto BBH systems.

In this study, we adopt the standard thin disk model with a constant
accretion rate for each of the triple disks associated with those BBH
systems (see section~\ref{sec:config}). It is possible, however, the
temperature structures of and radiation from these disks may deviate
from that of the thin disk model because of several factors,  BBH torques
\citep[e.g.,][]{Rafikov12}, accretion stream(s) connecting the
outer circum-binary disk with the inner disk(s), and the associated
(magnetohydrodynamical) shock(s) at both the inner edge of the outer
circumbinary disk and the outer edges of the inner disks.  Compared
with the thin disk, these disks might emit more power at the inner
edge of the circum-binary disk and at the outer edge of the disks
associated with the two BBH components.  This complication certainly
needs further investigation, though it may not significantly affect
the anomaly in the $r_{1/2}-\lambda$ relation found in this study
because of the existence of the gap.

\item Disk orientation. All the disk systems are assumed to be face on
in the above calculations. In reality, the background QSOs should not
be exactly face on. As the lensed QSOs are normally type 1 QSOs, they
have an inclination angle $i$, i.e., the angle between the disk normal
and the line of sight, roughly in the range of $\cos i \sim 0.5-1$
\citep[e.g.,][]{Krolik99}. Considering this orientation effect, the
surface brightness distributions for each disk in the non-face on
cases are elliptic-like. Comparing with the surface brightness
distributions shown in Figure~\ref{fig:f2} for the face-on cases,
those for non-face on cases shrink at one direction by $|\cos i|$
while maintaining the same at the other, orthogonal direction. This
change in the surface brightness distributions leads to corresponding
changes in the microlensing light curves. For illustration purpose, we
assume two BBH systems, which are similar to B5 and B8 except their
disks are non-face on. Our calculations show that the heights and
separations of the several magnification peaks in their light curves
are quite different from those shown in Figure~\ref{fig:f7}, mainly
because of the more significant asymmetric distribution of the surface
brightness in the non-face on cases. These differences may help us to
simultaneously constrain the disk orientation and the BBH nature. We
also note that the projected area of the emitting region for a disk
system, with an inclination angle of $i$, is a factor of $\cos i$
($\sim 0.5-1$) smaller than that of a face on system. This difference
only leads to a slightly underestimation of the source size but does
not affect the shape of the $r_{1/2}-\lambda$ relation and thus does
not introduce any anomaly in the $r_{1/2}-\lambda$ relation.

\item Reconstruction of BBH systems through their microlensing light
curves. In section~\ref{sec:disksize}, we estimate the
$r_{1/2}-\lambda$ relation from the mock light curves of BBH QSO
systems by adopting a single disk model. In principle, we may
reconstruct BBH systems once they are selected as BBH candidates based
on their light curves and $r_{1/2}-\lambda$ relations. By this
reconstruction, we may obtain strong constraints on the physical
parameters of the BBHs, such as $a_{\rm BBH}$, $M_{\bullet}$,
$q$, $f_{\rm E,1}$, and $f_{\rm E,2}$, etc. We defer the feasibility
study of the reconstruction of those BBH systems through their
microlensing light curves to a future study.

\end{itemize}

\section{Conclusions }\label{sec:discon}

In this paper, we propose a novel method to probe sub-parsec BBH QSOs
through the microlensing of lensed QSOs. If a QSO hosts a sub-parsec BBH
in its center, it is expected that the BBH is surrounded by a
circum-binary disk at outside region, each component of the BBH is
surrounded by a small accretion disk, and a gap opened by the small
component in between the circum-binary disk and the two small disks.
Assuming such a structure for some hypothetical BBH systems, we
generate mock microlensing light curves for BBH QSO systems with
typical physical parameters. We show that the microlensing light
curves of a BBH QSO system at some given bands can be significantly
different from those of a single MBH QSO system because of the
existence of the gap and the rotation of the BBH and its associated
small disks around the center of mass. We estimate the half-light
radius-wavelength relations from those mock light curves and find that
the obtained half-light radius-wavelength relations of BBH QSO systems
can be much flatter than that of single MBH QSO systems at a
wavelength range determined by the BBH parameters, such as the total
mass, mass ratio, separation, etc., which is primarily due to the
existence of the gap. Such a unique feature of BBH QSO systems can be
used to select and probe sub-parsec BBHs in a large number of lensed QSOs
to be discovered in the near future. 

Note that hints for the existence of BBHs in some QSO systems may be
revealed by their SEDs \citep[][see
section~\ref{subsec:diskmodel}]{GM12} and other spectral features (see
the introduction), but these features may be model dependent
and somewhat obscure due to complications in the emission processes.
The new method proposed in this paper can provide stronger physical
constraints on BBH systems and is less model dependent, as the
structure of the disk accretion associated with BBH systems can be
well resolved by microlensing. The signatures are particularly
striking when the rotation effects are present (see
Figures~\ref{fig:f7} and \ref{fig:f8}).

Many of the lensed QSOs may be single MBH systems, but some of them
might be accretion systems with BBHs in their centers. The occurrence
rate of BBHs in QSOs is an important quantity to determine the
efficiency of using microlensing to probe sub-parsec BBHs in lensed QSOs.
In principle, the BBH occurrence rate is determined by: (1) the active
level of the two MBHs within the gap; and (2) the residence time of
BBHs at the separation that has the most significant signals (i.e.,
$\sim 500-1000\rg$).  Most recent numerical simulations suggest
that significant accretion can be induced onto the two MBH components
although a low-density cavity (gap) at the center of the circum-binary
disk is maintained \citep[e.g.,][]{Hayasaki07, MM08, Roedig11,
Noble12, Shi12, DOrazio12}. In this paper, we assume that the level of
activities of sub-parsec BBH systems embedded in circum-binary disks are
more or less the same as that of normal bright QSOs, therefore, the
occurrence rate of BBH systems in QSOs is mainly determined by the BBH
residence time. For those BBHs with total mass in the range of $10^7
-10^9\msun$, mass ratio $q\sim 0.1-1$ and separation $\sim
500-1000 \rg$, the residence time is roughly in the range from a few
$10^5$ to a few $10^6$ yr, according to \citet[][see Figures 1 and 3
therein]{Haiman09}.  The lifetime of normal QSOs is estimated to be in
the range of a few times $10^7$ to $10^8$~yr from various
methods \citep[e.g.,][]{HH01, MW01, Jakobsen03, YL04, YL08, Marconi04,
WW06, Worseck07, GSP08, Shankar09, Shankar13, LY11}.  Thus the
occurrence rate of BBHs in lensed QSOs should be in the range of a few
thousandth to a few percent.

The total number of currently known lensed QSOs is $\sim 100$
(\citealt{Kochanek06}; http://www.cfa.harvard.edu/castles;
http://www-utap.phys.s.u-tokyo.ac.jp/\~sdss/sqls/).  According to the
BBH occurrence rate estimated above, at most, several of these lensed
QSOs could host BBH systems with separation $\la 1000\rg$, which may
be detectable through the microlensing event(s).  Ground-based
surveys, such as the Panoramic Survey Telescope and Rapid Response
System (Pan-STARRS; which is already operating), the planned Large
Synoptic Survey telescope (LSST) and Euclid, will monitor a large area
of sky in multi-bands from UV, optical to infrared for multiple times
over a long period (e.g., $10$~yr). It is expected that the
Pan-STARRS and the LSST will discover $\sim 3000-8000$ luminous QSOs
that are gravitational lensed into multiple images by foreground
galaxies \citep{Abell09, OM10}. It is possible that tens to one
hundred of these lensed QSOs contain BBHs in their centers according
to the BBH occurrence rate estimated above.  Long term monitoring of
these lensed QSOs by the Pan-STARRS, LSST, Euclid and other future
facilities may provide a powerful and efficient way to discover sub-parsec
BBHs unambiguously through the anomaly in the $r_{1/2}-\lambda$
relation estimated from the QSO light curves.

\acknowledgements 

This work was supported in part by the National Natural Science
Foundation of China under grant nos. 11103029 (CY), 11033001 (YL),
11273004 (QY), 11373031 (YL), and 11333003 (SM). CY, YL and SM thank
the financial support of National Astronomical Observatories, the
Chinese Academy of Sciences.

\appendix
\section{Root Mean Square histogram of the convolved maps and the
half-light radius}\label{subsec:rms}

\citet{Mortonson05} have demonstrated that the half-light radius of a
source can be well determined by the dispersion (rms) of the convolved
magnification histogram (see Figure 9 in \citealt{Mortonson05}). Note
that the magnification histogram represents the probability
distribution of the pixels in the convolved magnification map that
have a certain magnitude shift of the macroimage's flux with respect 
to its mean.  In order to better understand the results on the
$r_{1/2}-\lambda$ relations obtained above for the BBH systems through
the Bayesian fitting method, we also adopt the rms method given by
\citet{Mortonson05} to infer $r_{1/2}$, i.e., we first obtain the
dispersions (rms) of magnification histograms for all BBH systems and
their corresponding single MBH systems, and then use rms to infer
$r_{1/2}$. The magnification histogram of a BBH system does not depend
on whether we consider the rotation of the BBH and its associated disk
around the center of mass or not.

\begin{figure}
\centering
\includegraphics[scale=0.4,width=8.0cm,angle=-90]{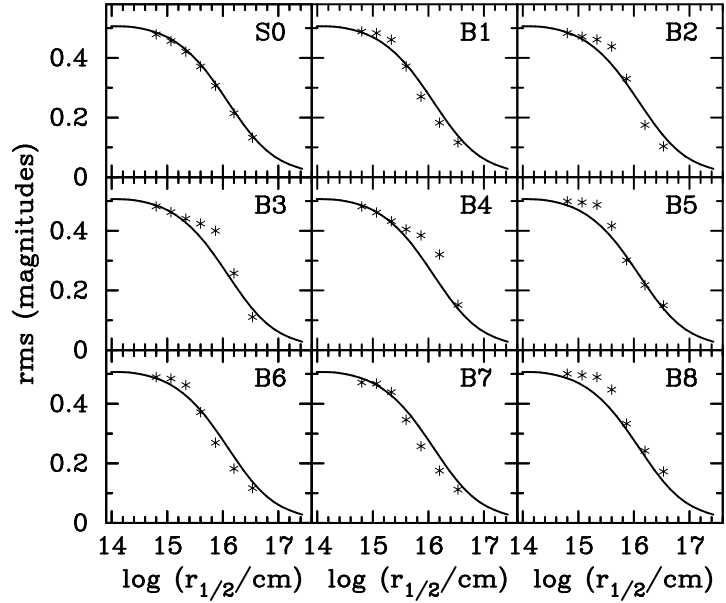}

\caption{RMS of the magnification histogram at different wavelengths
obtained from the magnification maps for the image A of the BBH
systems shown in Figure~\ref{fig:f4} (asterisks) and those of 
corresponding single MBH systems (solid line).  }

\label{fig:f16}
\end{figure}

\begin{figure}                                                
\centering                                                    
\includegraphics[scale=0.8,width=10.0cm,angle=-90]{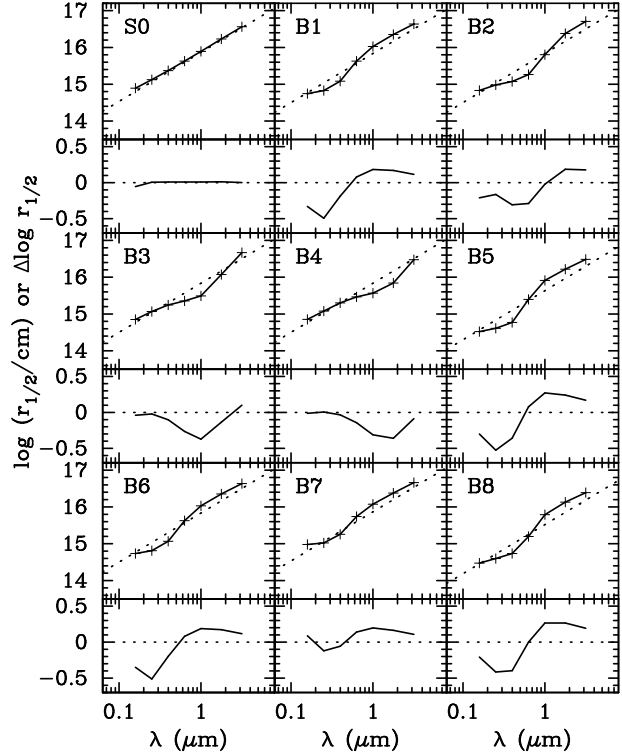}    
                                                              
\caption{Size-wavelength relation obtained from the rms estimation of
the image A for each system.  Panels with label of S0 to B8 show the
system S0 to B8 listed in Table~\ref{tab:t1}. In each panel with a
label, the plus symbols and solid curve represent the
$r_{1/2}-\lambda$ relation estimated from each single/triple disk
system, while the dotted line shows the expected $r_{1/2}-\lambda$
relation for a corresponding single disk system (see
equation~\ref{eq:rhalf}).  A small panel, below each labeled panel,
shows the difference between the expected $r_{1/2}-\lambda$ relation
for each BBH system and that for the corresponding single MBH system,
respectively. }

\label{fig:f17}
\end{figure}

Figure~\ref{fig:f16} shows the dispersion of the magnification
histogram (rms) for the BBH systems obtained from the convolved
magnification maps shown in Figure~\ref{fig:f4} and those of the
corresponding single MBH systems. As seen from Figure~\ref{fig:f16},
rms estimated for the BBH systems (asterisk symbols) obviously
deviates from that for its corresponding single disk systems (solid
line).  If we adopt the relation between rms and $r_{ 1/2}$ of the
single disk system to estimate the size of the emitting region of BBH
systems at a given wavelength, then the resulting $r_{1/2}$ may also
deviate from that expected for the single disk model (see
Equation~\ref{eq:rhalf}).  Figure~\ref{fig:f17} shows the resulting
relationship between the half-light radius and the wavelength for each
BBH system listed in Table~\ref{tab:t1} and its residual with respect 
to the relation expected from the corresponding single MBH system,
which are similar to that shown in Figure~\ref{fig:f7}.  The
consistency of these two results suggests that the magnification
histograms can be used to reveal the difference between the
$r_{1/2}-\lambda$ relation of a BBH system and that  of the
corresponding single MBH system, and this difference offers an easy
way to explore the parameter space of the BBH systems.                       

\end{document}